\documentclass[structabstract]{aa}  %printer format
\usepackage{graphicx}
\graphicspath{{figures/}}            % hide the figures in a subdirectory.
 \pdfoutput=1
\usepackage{txfonts}
\usepackage{natbib}
\bibpunct{(}{)}{,}{a}{,}{,}

\usepackage[pdfkeywords={},linktocpage=true,citecolor=red,linkcolor=blue,
            pdfpagemode={Fullscreen},bookmarks=false,bookmarksopen=false,
              colorlinks=true]{hyperref}

\begin{document}
 \title{Loss cone evolution and particle escape in collapsing magnetic trap models in solar flares}
 % \subtitle{I. Overviewing the $\kappa$-mechanism}

\author{Solmaz Eradat Oskoui \and Thomas Neukirch \and Keith James Grady
%\fnmsep\thanks{ }
         }

\institute{School of Mathematics and Statistics, University of St Andrews, St. Andrews KY16 9SS,UK
              \\
              \email{se11@st-andrews.ac.uk} \\
               \email{tn3@st-andrews.ac.uk}
            % \thanks{The university of heaven temporarily does not
             %        accept e-mails}
}

%\date{Received September 15, 1996; accepted March 16, 1997}

% \abstract{}{}{}{}{} 
% 5 {} token are mandatory
 
\abstract
   {Collapsing magnetic traps (CMTs) have been suggested as one possible mechanism responsible for the acceleration of high-energy particles during solar flares.
  An important question regarding the CMT acceleration mechanism is which particle orbits escape and which are trapped during the
  time evolution of a CMT. While some models predict the escape of the majority of particle orbits, other more sophisticated CMT models
  show that, in particular, the highest-energy particles remain trapped at all times. The exact prediction is not straightforward because
  both the loss cone angle and the particle orbit pitch angle evolve in time in a CMT.}
   {Our aim is to gain a better understanding of the conditions leading to either particle orbit escape or trapping in CMTs.
   }
   {
   We present a detailed investigation of the time evolution of particle orbit pitch angles in the CMT model of Giuliani and collaborators and compare this with the time
   evolution of the loss cone angle.
   The non-relativistic guiding centre approximation is used to calculate the particle orbits. We also use simplified models to corroborate the findings
   of the particle orbit calculations.}
   {We find that there is a critical initial pitch angle for each field line of a CMT that divides trapped and escaping particle orbits. This critical initial pitch angle is greater than
   the initial loss cone angle, but smaller than the asymptotic (final) loss cone angle for that field line. As the final loss cone angle in CMTs is larger than the initial loss cone
   angle, particle orbits with pitch angles that cross into the loss cone during their time evolution will escape whereas all other particle orbits are trapped. We find that in realistic
   CMT models, Fermi acceleration will only dominate in the initial phase of the CMT evolution and, in this case, can reduce the pitch angle, but that
   betatron acceleration will dominate for later stages of the CMT evolution leading to a systematic increase of the pitch angle.
   Whether a particle escapes or remains trapped depends critically on the relative importance of the two acceleration mechanism, which cannot be
   decoupled in more sophisticated CMT models.
   }
   {}


   \maketitle
%
%________________________________________________________________

\section{Introduction}

Collapsing magnetic traps (CMTs) have been suggested as one of the mechanisms that could contribute to
particle energisation in solar flares \citep[e.g.,][]{SomovKosugi}.
The basic idea behind CMTs is that charged particles will be trapped on the magnetic field lines below the 
reconnection region of a flare. 
The magnetic field will evolve into a lower energy state, resulting in (a) a shortening of field line length
and (b) an increase in the overall field strength.
Due to the vast difference in length and timescales between the particle motion and the magnetic field evolution,
the particle motion can be described to a large degree of accuracy by guiding centre theory. The conservation of 
a particle's magnetic moment and the bounce invariant \citep[e.g.,][]{Keith_grady_2012} give rise
to the possibility of an increase in the particle's kinetic energy by betatron acceleration and by first order Fermi acceleration
\citep[e.g.,][]{SomovKosugi,Boga_somov_2005}.

Studies of CMTs using models with varying degree of detail, focusing on different aspects of CMT physics have be 
carried out \citep[e.g.,][]{SomovKosugi}. In this paper, we investigate one very important aspect of particle motion in CMTs, namely what 
determines whether particles remain trapped or escape during the evolution of a CMT. Answering this question is important to
be able to assess whether the CMT mechanism can contribute to the energetic particle flux, causing hard X-ray emission from
the footpoints of flaring magnetic loops or whether it can be expected to contribute more to emission originating from higher up in
the corona \citep[e.g.,][]{Krucker_2008}.

At first sight, this may seem like a trivial question to answer because obviously as soon as a particle orbit moves
into the loss cone it will escape from the CMT. However, on second thought, things are not as simple as they seem for two reasons:
(a) the loss cone itself changes in time due to the changing magnetic field strength, and (b) the particle
pitch angle also evolves due to betatron and Fermi acceleration. Generally, one would expect the magnetic field
strength within a relaxing magnetic loop to increase. Assuming that the magnetic field strength is highest at the foot points and that
this field strength does not change substantially, the loss cone should generally open up during 
the evolution of a CMT. How much the loss cone opens depends on the magnetic field model \citep[see, e.g.,][for a model where the loss cone opens to
$90^\circ$]{Asch_2004}. Furthermore, whereas in investigations based on a number of simplifying assumptions \citep[e.g.,][]{Somov_2003,somov_2004,Boga_somov_2005}
some predictions can be made about, for example, the relative importance of betatron and Fermi acceleration and the resulting consequences. However, similar predictions are much harder to make for more detailed magnetic field models \citep[e.g.,][]{Giul2005, Kar_Barta_2006,Mino_2010,Keith_grady_2012} because the relative importance of betatron vs. Fermi acceleration: (i) will be different for particles with different initial conditions; (ii) may change with time during the evolution of the
CMT; and (iii) will depend on the details of the magnetic field model used. Furthermore, it is well-known that in time-dependent and curved magnetic fields the 
two mechanisms are closely linked \citep[see, e.g.,][]{northrop}.

The aim of this paper is to present a detailed investigation of particle escape and trapping for the CMT model of \citet{Giul2005}. For the purpose of this paper, we ignore pitch angle scattering by Coulomb collisions, wave-particle interaction, or turbulence. Pitch angle scattering will, of course, change the results, but we regard
the present investigation as a benchmark with which possible future investigations, including the effects of pitch angle scattering, can be compared.
In Sect. \ref{sec:model},  we give a brief outline of the CMT model of \citet{Giul2005} and summarise some of the results of \citet{Keith_grady_2012} that are
relevant to this paper. In Sect. \ref{sec:pitch}, we investigate how the loss cone evolves in our CMT model and how the pitch angle of typical
particle orbits change with time, and we try to give an explanation of our results on the basis of some simplified models in Sect. \ref{sec:simplemodel}.
We present a summary of our results in Sect. \ref{sec:conclusions}

%__________________________________________________________________

\section{Brief model overview}
\label{sec:model}

\citet{Giul2005} used the kinematic MHD equations,
%Following a frame work developed by \cite{Giul2005} for the translationally invariant 2.5D CMT model, the basic theory uses the ideal kinematic MHD equations,
\begin{eqnarray}
{\mathbf{E}} + {\mathbf{v}} \times {\mathbf{B}} &=& {\mathbf{0}}, \label{ohms_law}\\
\frac{\partial {\mathbf{B}}}{\partial t} &=& - \nabla \times {\mathbf{E}}, \label{Faraday}\\
\nabla \cdot{\mathbf{B}} &=&0 ,\label{divBzero}
\end{eqnarray}
to develop a general framework for analytical CMT models, describing the evolution of the magnetic field for a 
given flow velocity, $\mathbf{v}({\mathbf{x}},t)$ under the assumption that the magnetic field is translationally 
invariant in the $z$-direction and the flow velocity has vanishing $v_z$ \citep[for an extension to fully three-dimensional
model, see][]{grady:TN09}. It is assumed that the $x$ and $z$ coordinates run parallel to the solar surface and
$y$ represents the height above the solar surface.

Under the assumption of translational invariance, one can use a flux function $A(x,y)$ to write the magnetic field in the
form
\begin{equation}
{\mathbf{B}} = \nabla A \times {\mathbf{e}}_z + B_{z}{\mathbf{e}}_z, \label{B_field}
\end{equation}
which automatically satisfies Eq. (\ref{divBzero}). The other two equations can be written as
\begin{eqnarray}
\frac{dA}{dt} = \frac{\partial A}{\partial t} + {\mathbf{v}}.\nabla A &=& 0, \label{flux_function} \\
\frac{\partial B_z}{\partial t} + \nabla . (B_z{\mathbf{v}}) &=& 0, \label{Bz_eqn}
\end{eqnarray}
with the electric field given by
\begin{equation}
{\mathbf{E}} = - \frac{\partial {\mathbf{A}}}{\partial t}. \label{Efield}
\end{equation}
Following \citet{Giul2005} we will assume that $B_z$ and $v_z$ vanish.

A CMT model is then defined by choosing a form for the flux function $A_0(x,y)=A(x,y, t_0)$ at a fixed time $t_0$,
and by specifying a flow field $\mathbf{v}(x,y,t)$. In the theory of \citet{Giul2005}, instead of defining the flow field
directly, it is given implicitly by choosing a time-dependent transformation between Lagrangian and Eulerian coordinates.
The advantage is that one can then immediately solve Eq. (\ref{flux_function}) using $A_0(x,y)$ as the initial or, as was the choice of 
\citet{Giul2005}, the final condition.

\begin{figure}
\resizebox{\hsize}{!}{\includegraphics[trim= 0cm 1.2cm 0cm 2cm, clip=true,width=6cm]{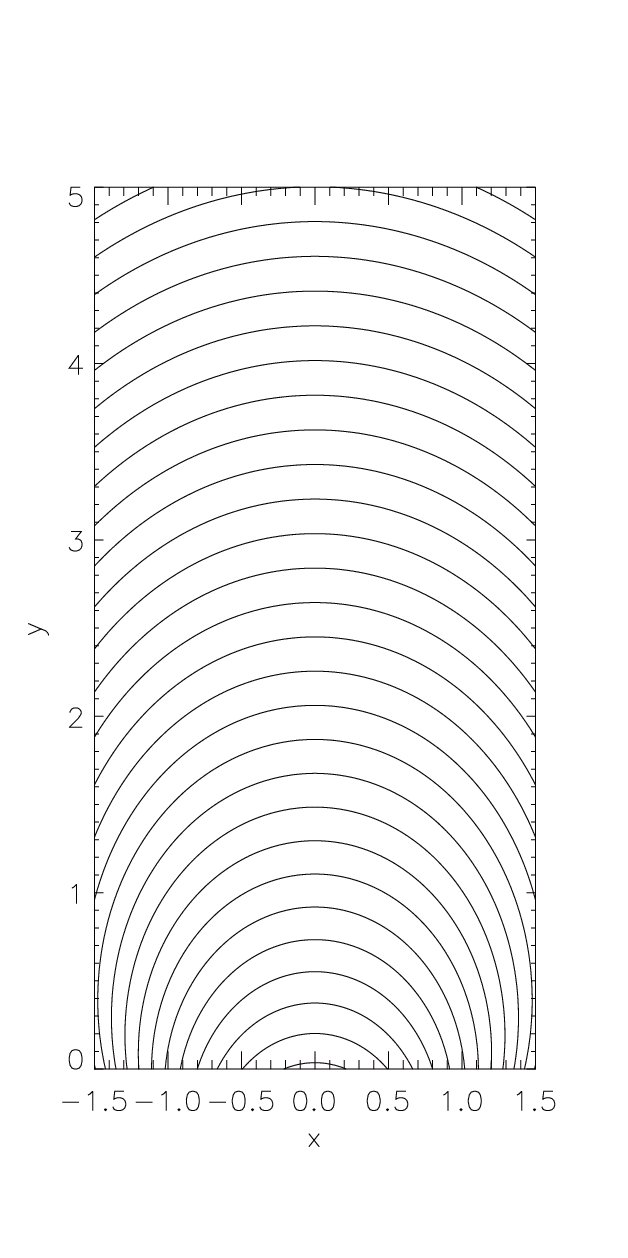}}
\caption{Field line plot of the asymptotic state ($t\to \infty$) of the magnetic field of the CMT model used in this paper.}
\label{fig:fluxfinal}
\end{figure}

In this paper, we use the same magnetic field model as \citet{Giul2005} and \citet{Keith_grady_2012}, which is given by
\begin{equation}
A_0 =  c_1\left[ \arctan\left(\frac{y_0+d_1}{x_0+w}   \right) - \arctan\left(\frac{y_0+d_2}{x_0-w}   \right)  \right].
\label{defA0}
\end{equation}
This flux function represents a potential magnetic field loop at time $t_0$ (we assume that $t_0 \to \infty$) created by two line sources of
strength $c_1$ (one of positive and one of negative polarity) located at the positions $(-w,-d_1)$ and $(w,-d_2)$ below the lower
boundary. All lengths here are scaled to a fundamental length scale $L$. Following \citet{Giul2005}, we choose $w=0.5$ and
$d_1=d_2=1.0$, creating a symmetric magnetic loop (Fig. \ref{fig:fluxfinal}).

The CMT model is completed by choosing the time-dependent transformation between Eulerian and Lagrangian coordinates.
Again, we choose the same transformation as in \citet{Giul2005} and \citet{Keith_grady_2012}, namely
\begin{eqnarray}
 x_0 &=& x,\label{x_infinity}\\
 y_0 &=& (at)^{b} \ln\left[ 1 + \frac{y}{(at)^{b}}\right] \left\lbrace \frac{1+ \tanh\left[ (y - L_v / L ) a_1\right]}{2}\right\rbrace \nonumber \\ 
           && + \left\lbrace \frac{1- \tanh\left[ (y - L_v / L ) a_1\right]}{2}\right\rbrace y  \label{y_infinity}.
\end{eqnarray}
The parameter $L_v$ is the characteristic height above which the magnetic field  is stretched by the transformation. Below this height the magnetic field is largely unchanged. 
The parameter $a_1$ determines the scale over which this transition from an unstretched to a stretched field takes place, whereas $a$ and $b$ are parameters that are related to the timescale of the evolution of the CMT.  We use the same values for the parameters as in the previous papers, 
i.e., $a = 0.4$, $b = 1.0$, $L_v/L = 1.0$, and $a_1 = 0.9$.

%_______________________________________________________________________________
\section{Evolution of pitch angle $\theta$ and loss cone $\alpha$ for non-relativistic particle orbits} 

\label{sec:pitch}

%non_rel_5_6_keV_alpha_fixed
%______________________________________________________________
 \begin{figure}
 \resizebox{\hsize}{!}{
   \includegraphics[width=6.75cm]{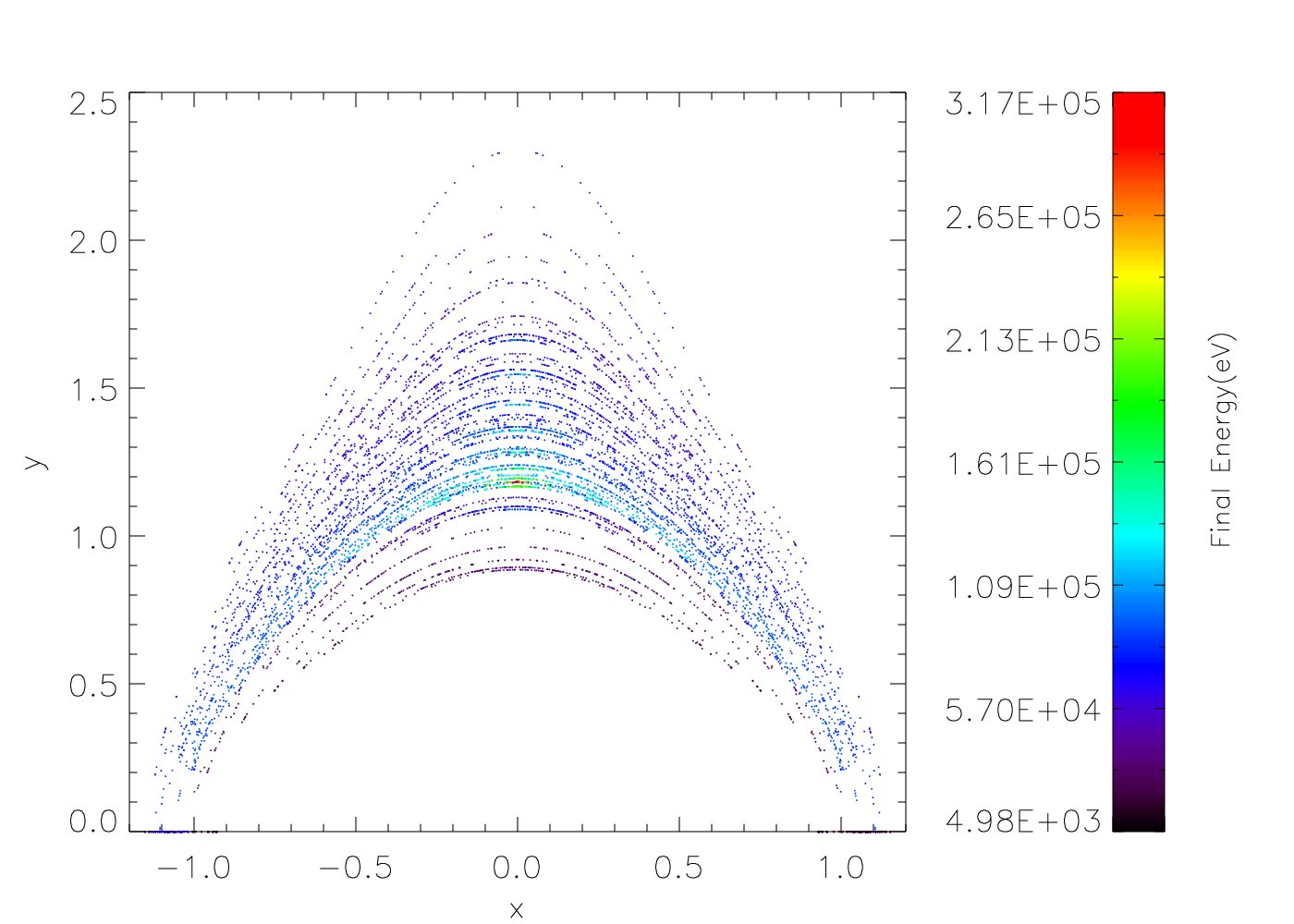}}
      \caption{ Graph of final position with corresponding energy. The colour
		bar gives the final energy of the trapped particles. The highest
		energies are found to be trapped in the middle of the trap.
}
         \label{FigfinalEpos}
   \end{figure}

Due to the vast difference between the Larmor frequency and radius of charged particle orbits and the time and length scales of the CMT model,
one can safely make use of the guiding centre approximation for calculating particle orbits \citep[see, e.g.,][]{Giul2005}. It turns out that the 
$\mathbf{E}\times\mathbf{B}$-drift is by far the dominating drift and that therefore the guiding centre remain on (or very close to) the same field for all times
\citep[see, e.g.,][for a detailed discussion]{grady2012}.

%ideas for our investigation************
We first briefly summarise the findings of 
\cite{Keith_grady_2012}, who like \citet{Giul2005} stopped the integration of particle orbits after a finite time (corresponding to 95 seconds in their normalisation).
\cite{Keith_grady_2012} found that for most initial conditions the particle orbits remain trapped during this time. 
They also found that, not surprisingly, different initial positions $(x,y)$, initial energies 
$E$ and initial pitch angles $\theta$ have an effect on the position of the mirror points, the energy gain of the particle orbits, and on whether they remain trapped or escape. 

In particular, they found that the particle orbits that gain most energy during the trap collapse have initial pitch angles $\theta$ close to $90^{\circ}$ and initial positions
in a weak magnetic field region in the middle of the trap. These particle orbits with the largest energy gain remained trapped during the collapse and due to their pitch angle
staying close to $90^{\circ}$ have mirror points very close to the centre of the trap. \cite{Keith_grady_2012} argue that these particle orbits are energised mainly by the betatron
mechanism. Other particle orbits with initial pitch angles closer to  $0^{\circ}$ (or $180^{\circ}$) seem to be energised by the Fermi mechanism at the beginning, but
as already pointed out by \citet{Giul2005} and corroborated by \cite{Keith_grady_2012}, these particle orbits gain energy when passing through the centre of the trap. At later stages, these particle orbits also seem to be undergoing mainly betatron acceleration. We will make use of this result later in this paper.
To illustrate these findings we show the positions and energies (colour coded) at the end of the nominal collapse time of 95 seconds for  a number of initial conditions in 
Fig. \ref{FigfinalEpos}. Particle orbits that have crossed the lower boundary (i.e., escaped) are represented by the dots on the $x$-axis of the plot. 

Before we start a more detailed  investigation of particle trapping and escape in our CMT model,
we recall a number of basic definitions that we make use of throughout the paper.
We already mentioned the pitch angle of a particle orbit, which is defined as
\begin{eqnarray}
 \theta = \arccos \left( \frac{v_{\parallel}}{v}\right),
 \label{pitchangledef}
\end{eqnarray}
where $v_\parallel$ is the velocity of the particle along the field line and $v$ is the total velocity of the particle. We define the mirror ratio as 
\begin{eqnarray}
 R(t) = \frac{B_{fp}}{B(t)},
 \label{mirrorratiodef}
\end{eqnarray}
where $B_{fp}$ is the foot point field strength of a particular field line and $B(t)$ the field strength on the same field line at $x=0$.
The field strength at the foot points of the field lines basically remains constant over time because the transformation given
in Eqs (\ref{x_infinity}) and (\ref{y_infinity}) ensures that the $B_y$ component of the magnetic field on the lower boundary $y=0$ does not change, 
although the $B_x$ component could change in time. This change is so small, however, that one can regard the absolute value of the magnetic field strength
at the footpoints as constant. 
The mirror ratio $R(t)$ is related to the loss cone angle by
\begin{eqnarray}
 \alpha(t) =  \arcsin \left(\frac{1}{\sqrt{R(t)}} \right).
 \label{lossconedef}
\end{eqnarray}
We emphasize that both the mirror ratio and the loss cone angle not only vary in time due to the time evolution of the magnetic field, but also from 
field line to field line, but we have suppressed the spatial dependence in the definitions above for ease of notation.

%______________________________________
%par_orbit_87_160.pdf work out in 
%Fortran:non_rel_0_4 .2_5.5_alpha_87.3
%idl: .r plot_compare
%________________________________________

\begin{figure}
\centerline{
\resizebox{\hsize}{!}{\includegraphics[width=6.75cm]{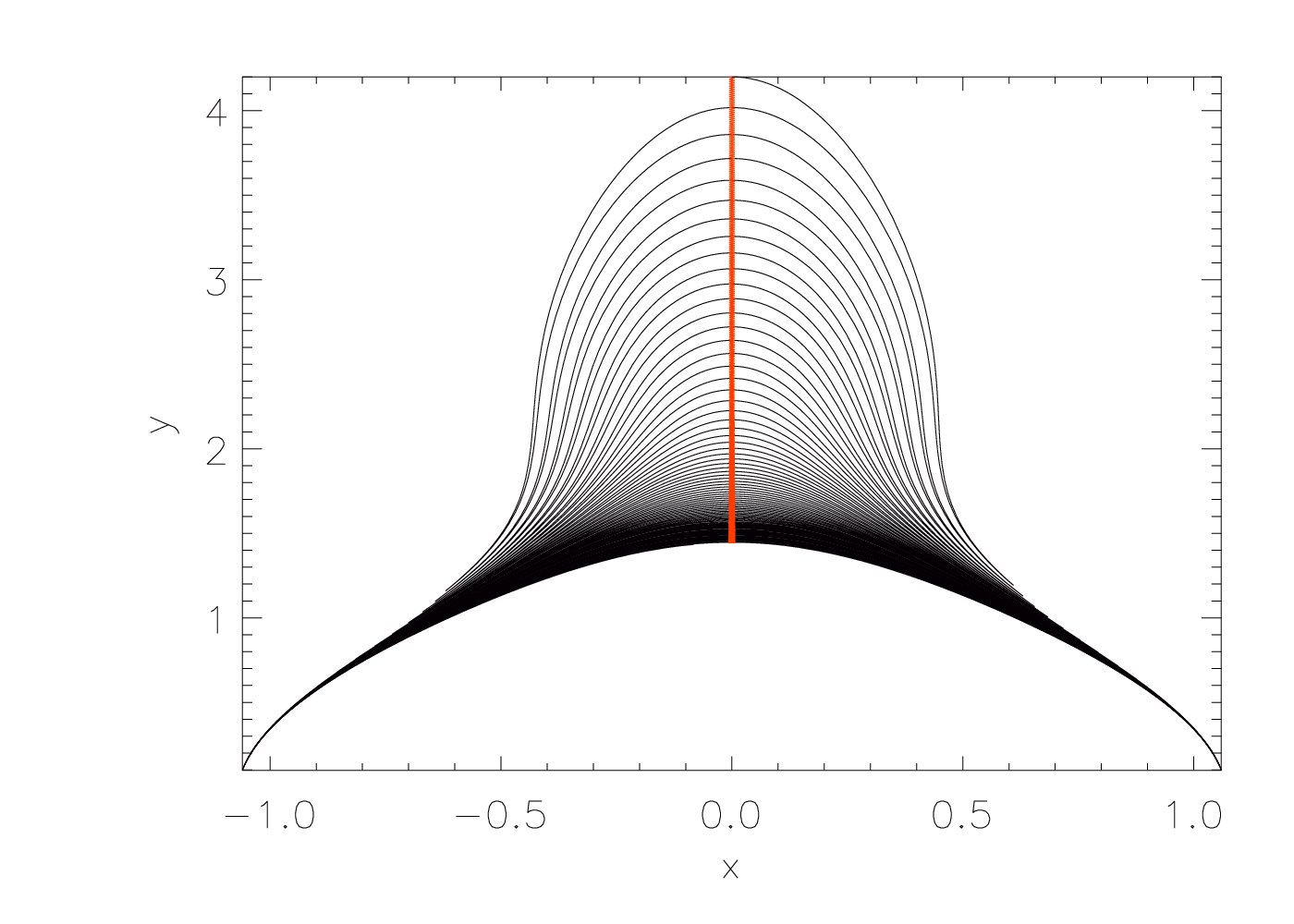}}}
\caption{Two illustrative particle orbits starting at the same initial position $x=0$, $y=4.2$ and with the same initial energy ($5.5$ keV),
but different initial pitch angles. Orbit 1 (red) has an initial pitch angle of $87.3^\circ$ (i.e. close to $90^\circ$) and hence stays close
to the middle of the CMT at all times, whereas orbit 2 (black) has an initial pitch angle of $160.4^\circ$ and has mirror points close to the lower
boundary.}
\label{fig:particleorbits}
\end{figure}

We start our investigation by looking in more detail at two particle orbits starting at the same initial position ($x=0$, $y=4.2$) and the same initial energy ($5.5$ keV),
but with different initial pitch angles of $87.3^\circ$ (orbit 1) and $160.4^\circ$ (orbit 2). 
These initial conditions are representative of the typical behaviour of
particle orbits with an initial pitch angle close to $90^\circ$, which have very little movement along the field lines, and particle orbits which 
at the initial time have a much larger velocity component 
parallel (or in this case anti-parallel) to the magnetic field. The two sets of initial conditions chosen here are very similar (albeit not
identical) to the two examples of orbits discussed in \citet{Keith_grady_2012}, and thus the orbits {(shown in Fig. \ref{fig:particleorbits})} are very
similar to the orbits discussed in their paper. As one can clearly see in Fig. \ref{fig:particleorbits}, orbit 1 (red) remains confined to the middle of the trap due to the fact
that the initial pitch angle is close to $90^\circ$, whereas orbit 2 (black) has mirror points very close to the bottom boundary, but does not escape during the time of the 
calculation. We also point out that there is very little change in the position of the mirror points over time, which is  consistent with previous findings
\citep[][]{Keith_grady_2012}.

\begin{figure*}
\centerline{
\includegraphics[width=0.497\hsize]{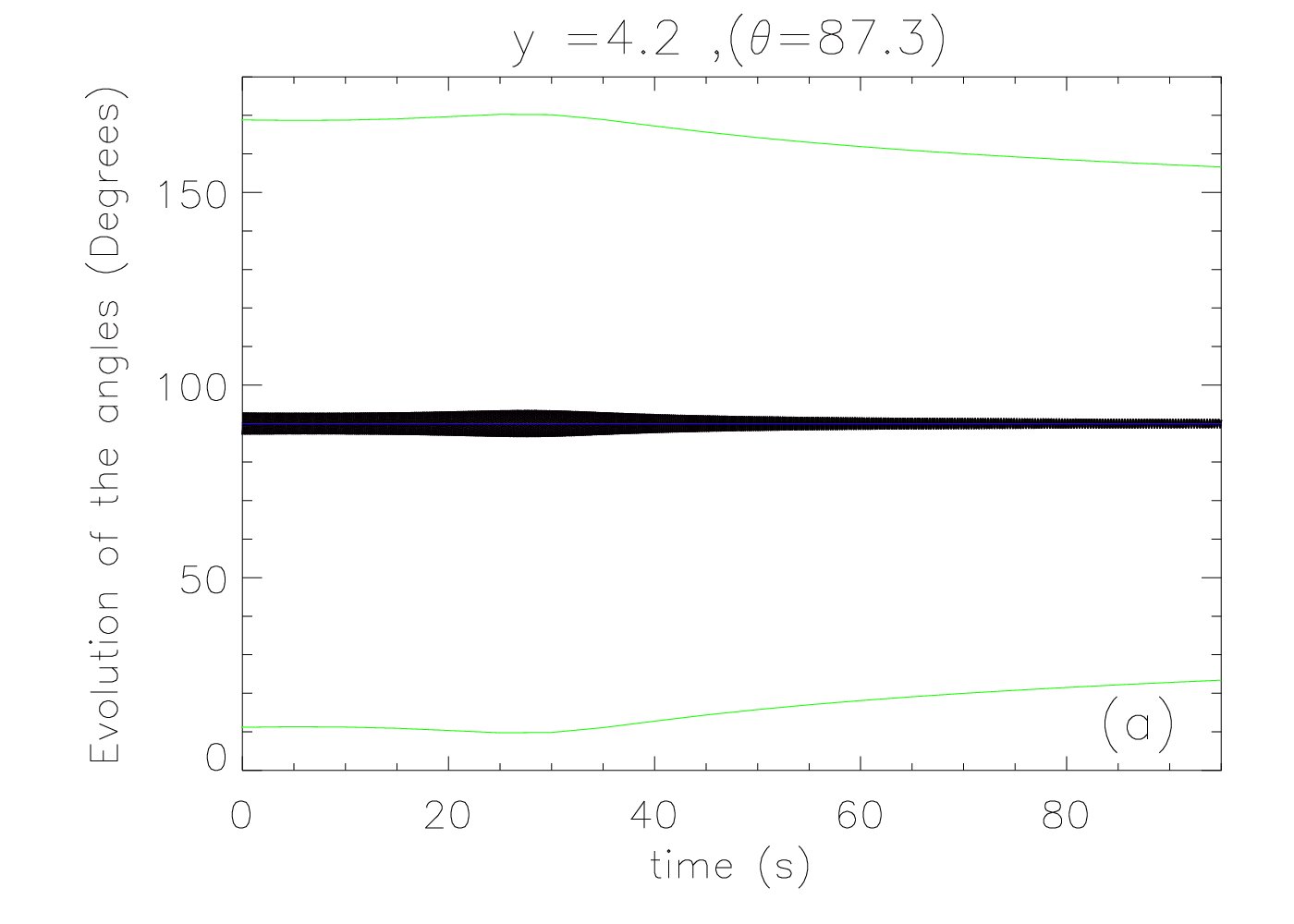}
\includegraphics[width=0.497\hsize]{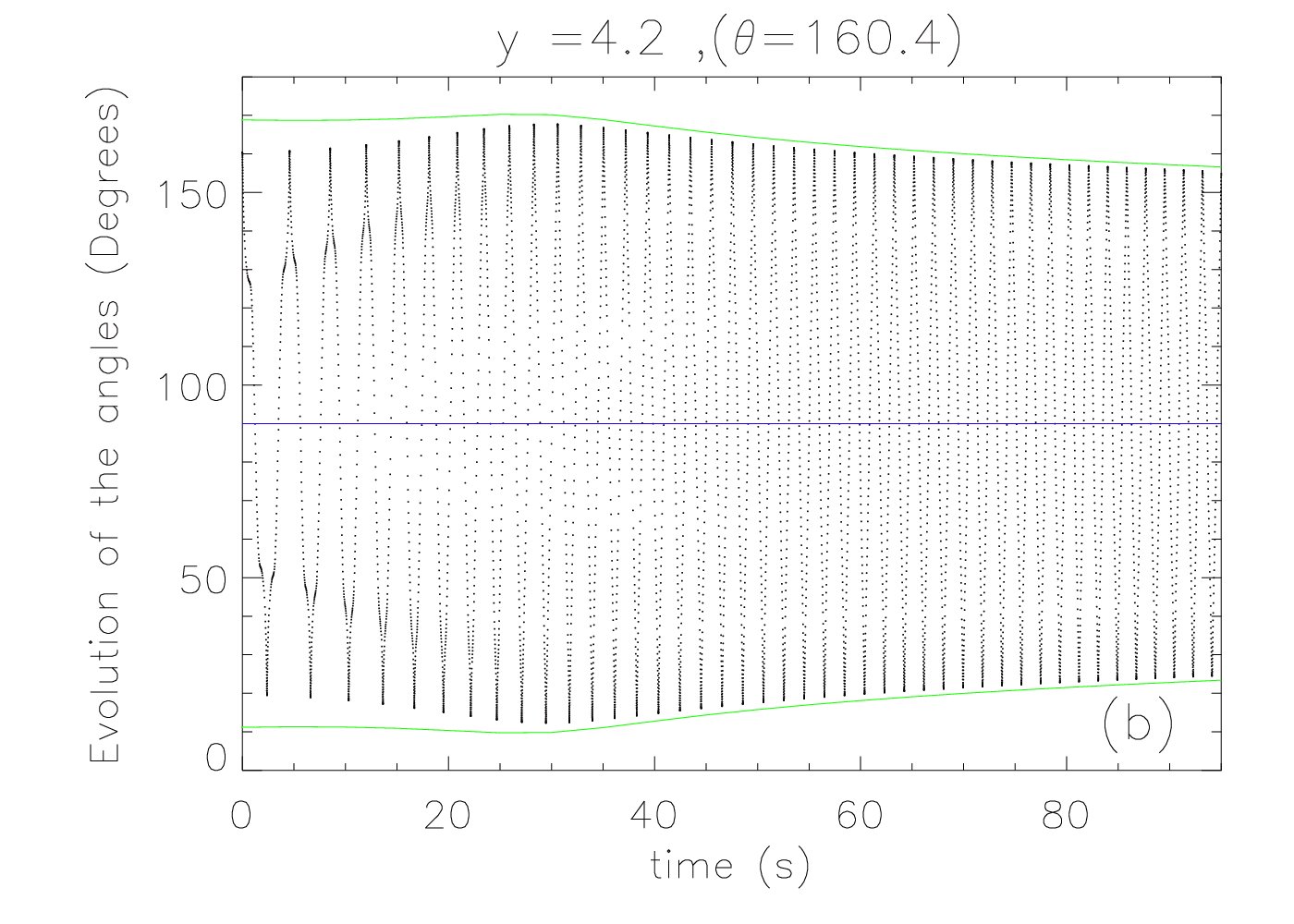}}
\label{fig:pitchang1and2}
  \caption{{The evolution of the particle orbit pitch angle $\theta$ (black dotted line) is compared with the evolution of the loss cone angle $\alpha$ (green lines) for
  particle orbits 1 (initial pitch angle $\theta= 87.3^{\circ}$) and 2 (initial pitch angle $\theta= 160.4^{\circ}$). 
The blue line marks the $90^{\circ}$ point (see main text for discussion).
%In both plots we show the loss cone angle $\alpha(t)$ as defined in Eq. (\ref{lossconedef}) and $180^\circ - \alpha(t)$. 
%As expected, the pitch angle for orbit 1 oscillates around the $90^\circ$ line with a slightly decreasing amplitude.
%Orbit 2 seems to approach the loss cone angle and then follow it without crossing into the loss cone within the time of the calculation.
}}
  \label{Figlosspitchevo}
\end{figure*}

To analyse the situation further, we have calculated the time evolution of the loss cone angle, as defined in Eq. (\ref{lossconedef}), for the field line on which
the orbits start and followed its time evolution to compare it with the time evolution of the pitch angles of the two particle orbits. The results are shown in 
Fig. \ref{Figlosspitchevo}. In the figures, the time evolution of the loss cone angle $\alpha(t)$ for the magnetic field line passing through the initial positions of the particle orbits at the 
initial time is shown in green (towards the bottom of the plots). We also show the angle $180^\circ - \alpha(t)$ (green line towards the top of the plots) because particle orbits
will escape from the trap if their pitch angle becomes either less than $\alpha(t)$ or larger than $180^\circ - \alpha(t)$. Obviously, the graph of $180^\circ - \alpha(t)$ here
is just a mirror image of the graph of $\alpha(t)$ when mirrored at the line $90^\circ$ (shown in blue on the plots) 
because the CMT we consider here has a magnetic field that is symmetric with respect to $x=0$. The initial value of the loss cone angle $\alpha(t)$ for this particular
magnetic field line is $11.8^\circ$ and its value at the end of the calculation is $23.4^\circ$. The asymptotic value of the loss cone angle in the limit $t\to\infty$ is, however,
much larger, namely $44.8^\circ$. As expected, the loss cone angle generally increases with time because the magnetic field strength at the apex of the magnetic field
line at $x=0$ increases with time. However, one can see from the plots that up to a time of roughly $30$ seconds the loss cone angle is decreasing. This is a particular feature of the
CMT model of \citet{Giul2005}, which in the initial stages of its time evolution has a region of very weak magnetic field through which the field lines have to move first, 
before the field strength starts to increase again. This is the reason for the initial dip in the graph of the loss cone angle. The asymptotic value of the loss cone angle does not approach $90^\circ$ because we have an 
inhomogeneous asymptotic magnetic field with a larger field strength at the foot points compared to the highest points along field lines, and thus particle
orbits can remain trapped in this CMT model \citep[e.g., contrary to the model of][]{Asch_2004}.
We emphasize again that the values of the loss cone angle at all times are different for different initial positions, but that the qualitative behaviour is the same. 

The time evolution of the pitch angle of particle orbits 1 and 2 is shown by the black dotted lines in Fig. \ref{Figlosspitchevo} (a) and (b), respectively. One can see in Fig. \ref{Figlosspitchevo} (a) the pitch angle of particle orbit 1 always remains very close to the blue $90^\circ$ angle. Although the oscillatory behaviour itself is very hard to see in the plot, the amplitude of the pitch angle oscillation is clearly decreasing as time progresses.
This can be easily understood by recalling that  \citet{Keith_grady_2012} showed that particle orbits of this type do not experience any significant changes in their
parallel energy, whereas they gain considerable amounts of perpendicular energy due to betatron acceleration. From Eq. (\ref{pitchangledef}), it is clear that if $v_\parallel$ remains 
roughly constant while $v$ increases due to the increase in the perpendicular velocity that $\theta$ will tend towards $90^\circ$.

The more interesting of the two particle orbits shown is orbit 2 (Fig. \ref{Figlosspitchevo} b), because its pitch angle is much closer to the loss cone angle and thus close to escaping from the CMT. Before we start a detailed discussion of the features seen in the plot, we remark that the maxima and minima of the pitch angle curve correspond to the
times when the particle orbit passes through the apex of the field line, whereas the mirror points correspond to the crossings of the $90^\circ$ line (blue). As already stated above, 
this particle orbit has an initial pitch angle of $160.4^\circ$ and therefore a much larger initial $v_\parallel$ than orbit 1. 
We see that in the first roughly $30$ seconds the maximum (minimum) values of the pitch angle curve increasing (decreasing) {are} approaching the green loss cone angle curve. After this time the pitch angle maxima (minima) start decreasing (increasing) and follow the same trend as seen in the curves representing the loss cone angle and its mirror image, while still very slowly approaching those curves.
This general trend seen in the pitch angle behaviour can be explained in a similar way to the behaviour of particle orbit 1 by  the increase in magnetic field strength
at the apex of the field line and the corresponding average gain in perpendicular energy due to betatron acceleration, with the main difference being that orbit 1 always 
remains close to the field line apex, whereas orbit 2 makes lengthy excursions down the field line almost to the lower boundary. The orbit does, however, not cross the
lower boundary before the calculation was stopped.

 \begin{figure*}
  \centerline{
   \includegraphics[width=0.45\hsize]{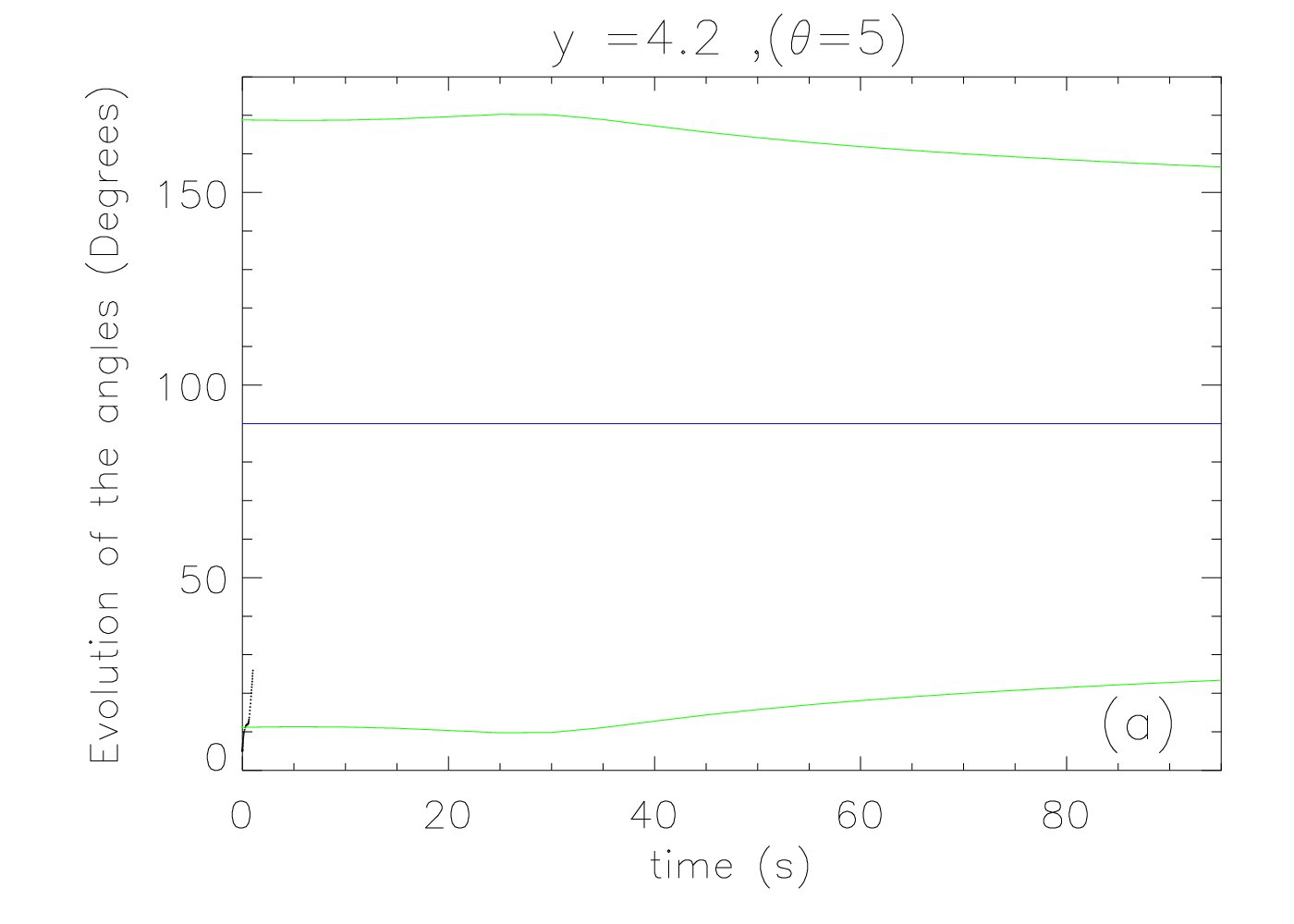}
   \includegraphics[width=0.45\hsize]{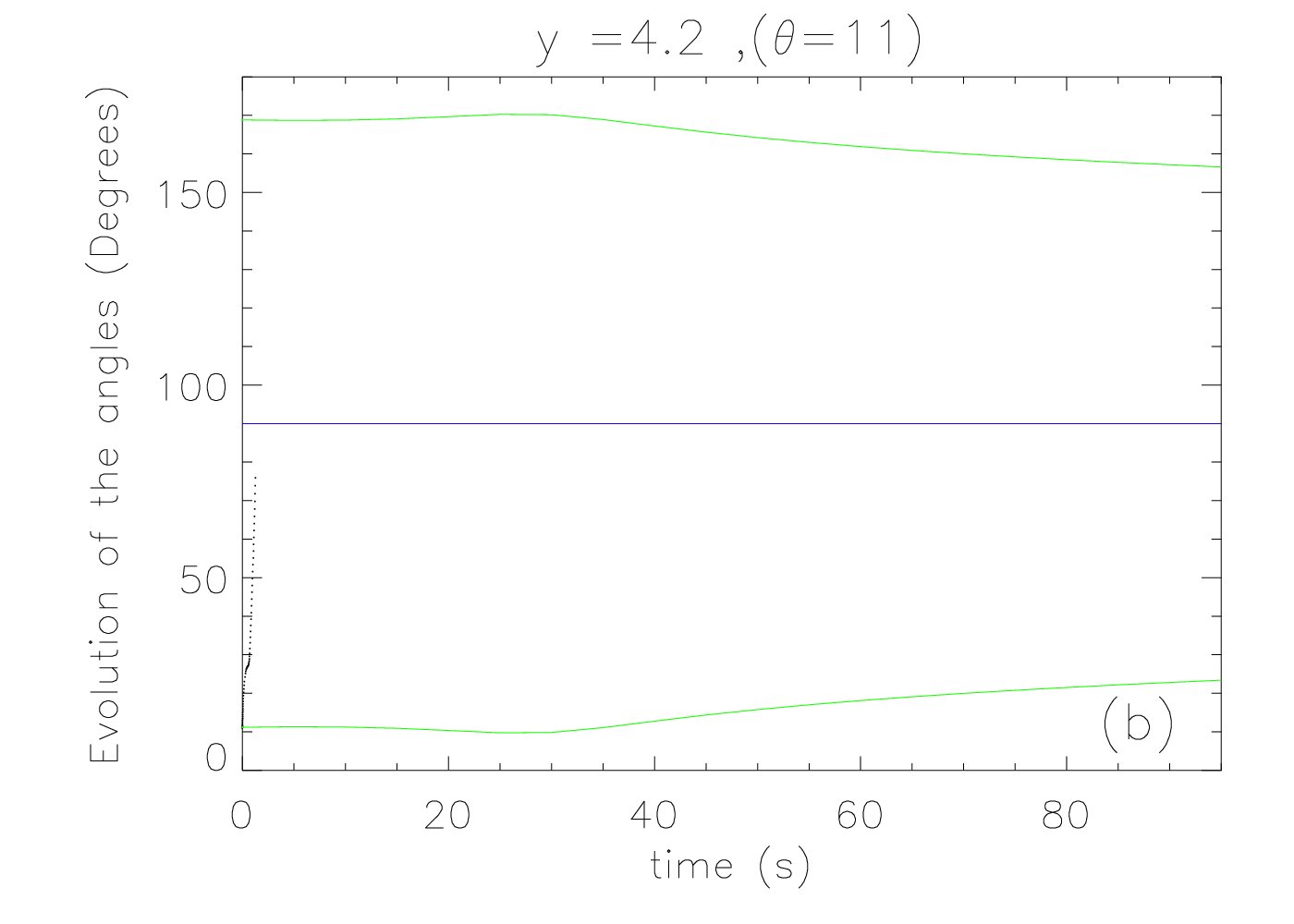}}

\centerline{
   \includegraphics[width=0.45\hsize]{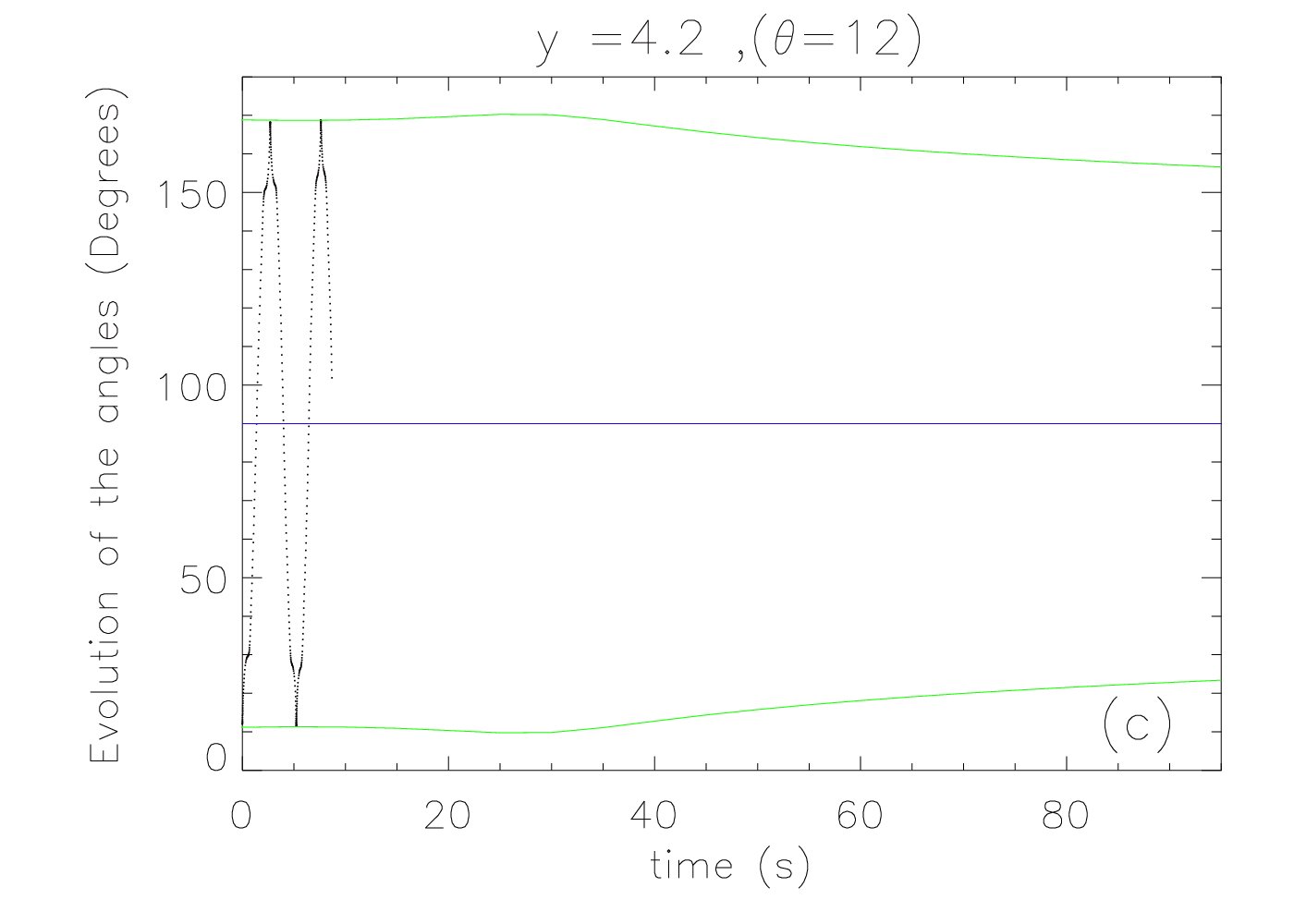}
   \includegraphics[width=0.45\hsize]{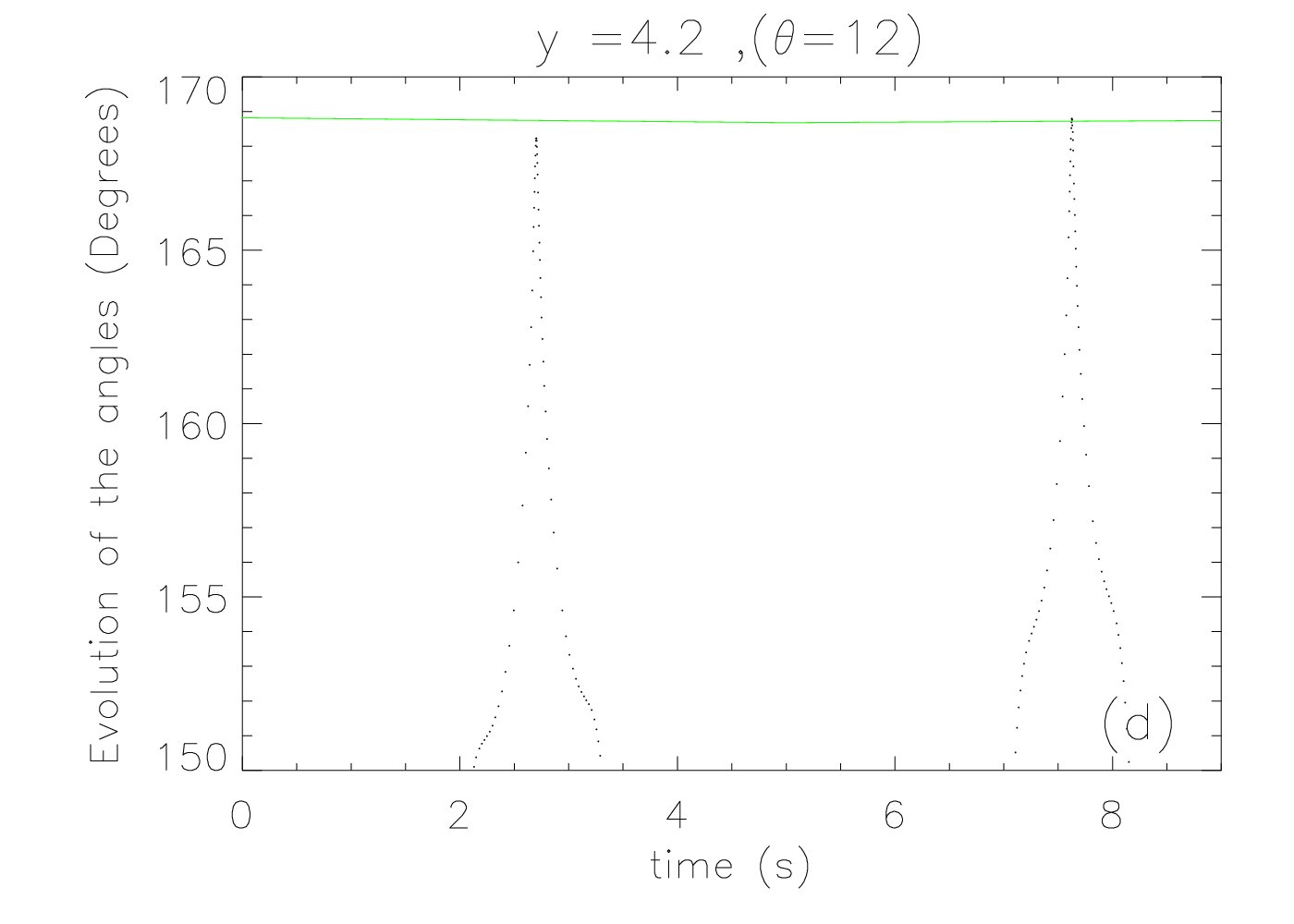}}

 \centerline{
   \includegraphics[width=0.45\hsize]{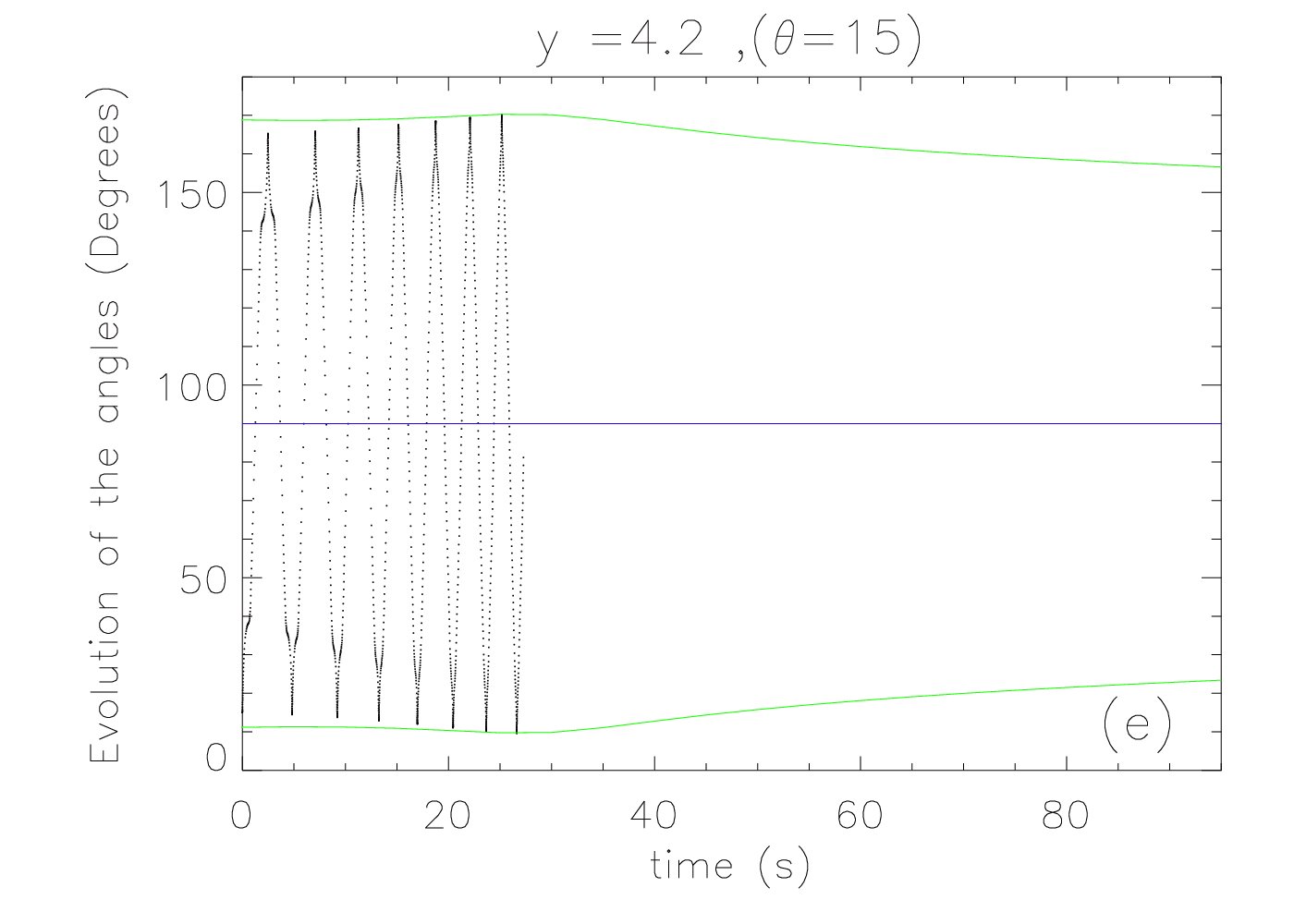}
   \includegraphics[width=0.45\hsize]{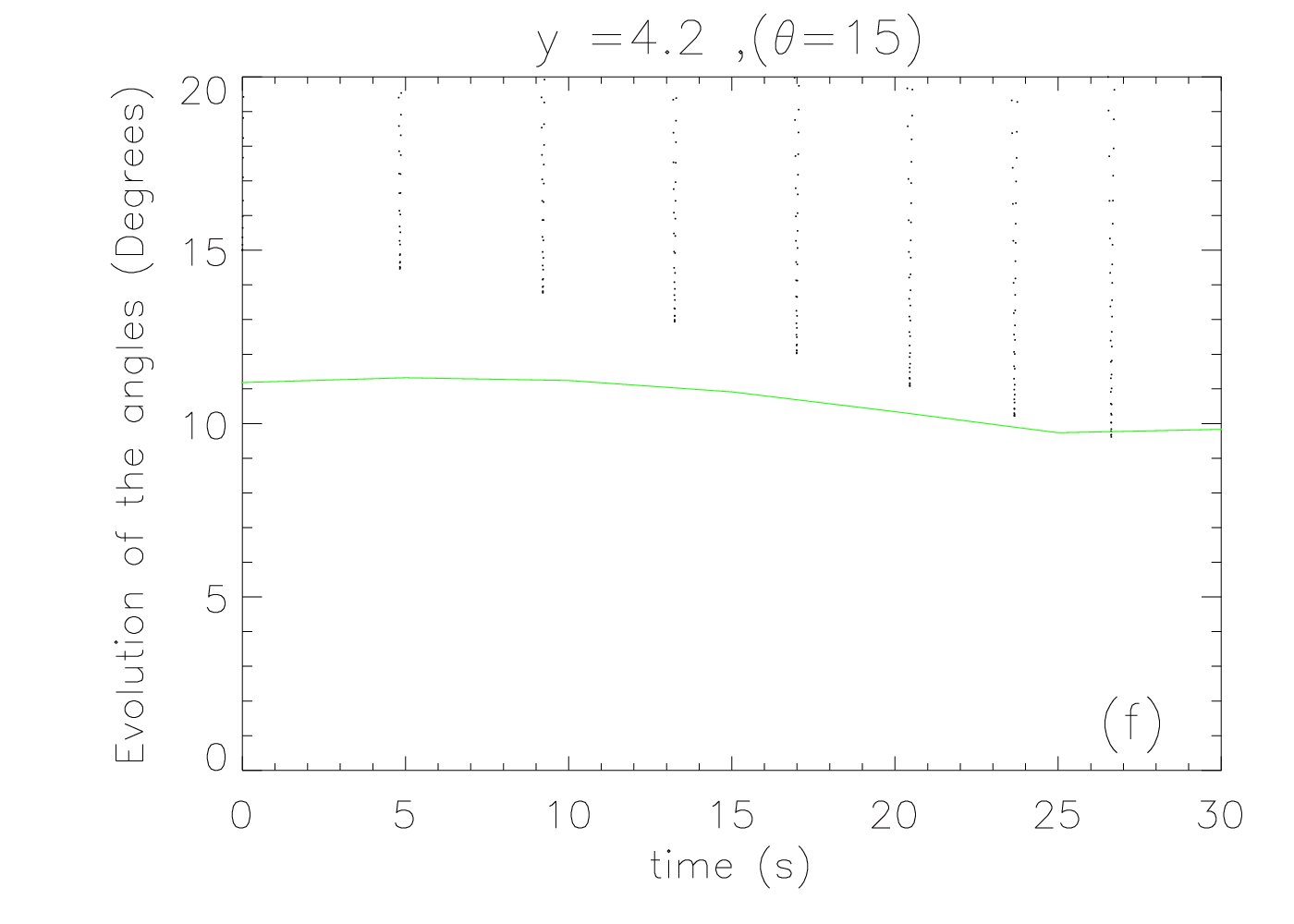}}

 \centerline{
   \includegraphics[width=0.45\hsize]{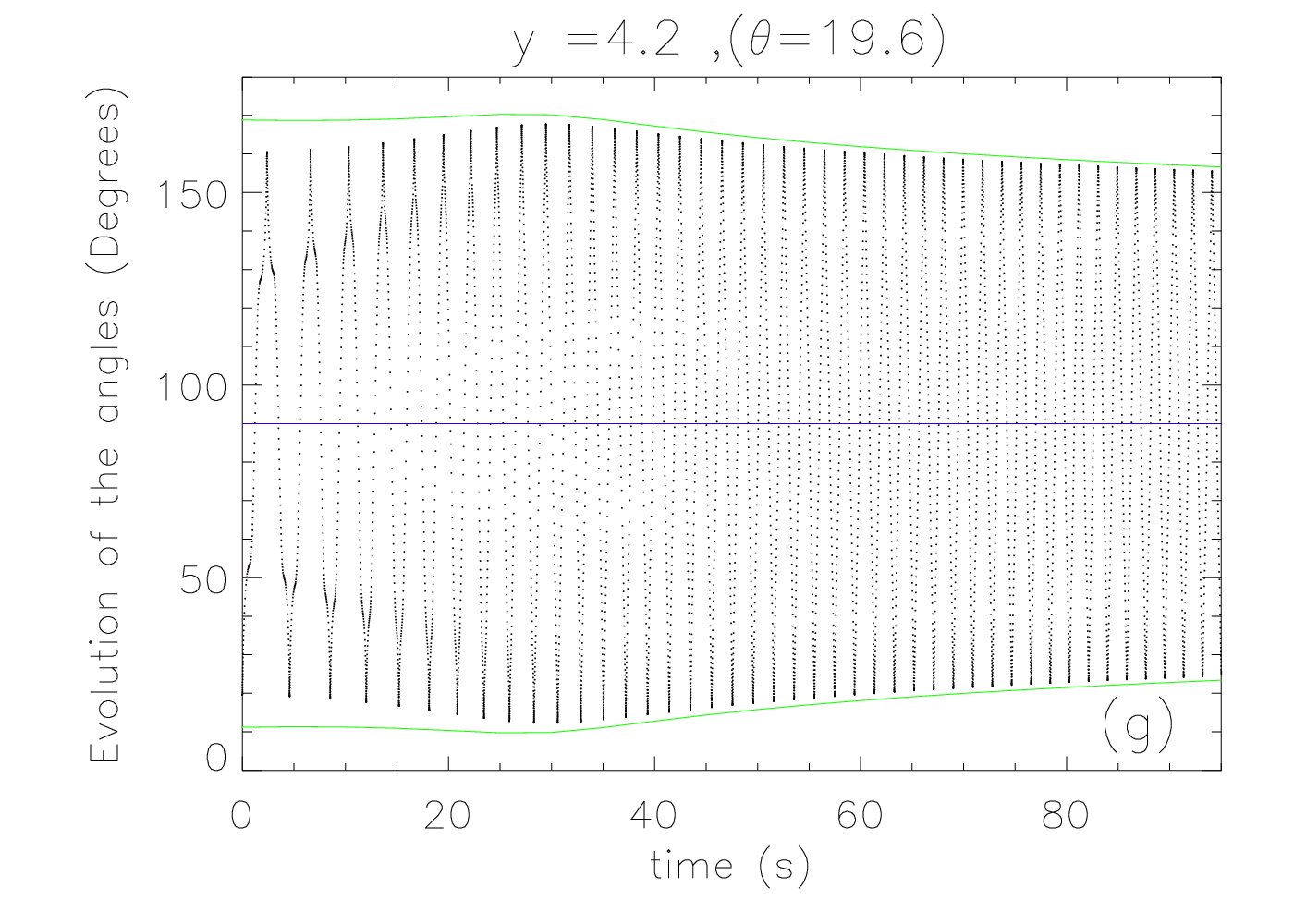}
   \includegraphics[width=0.45\hsize]{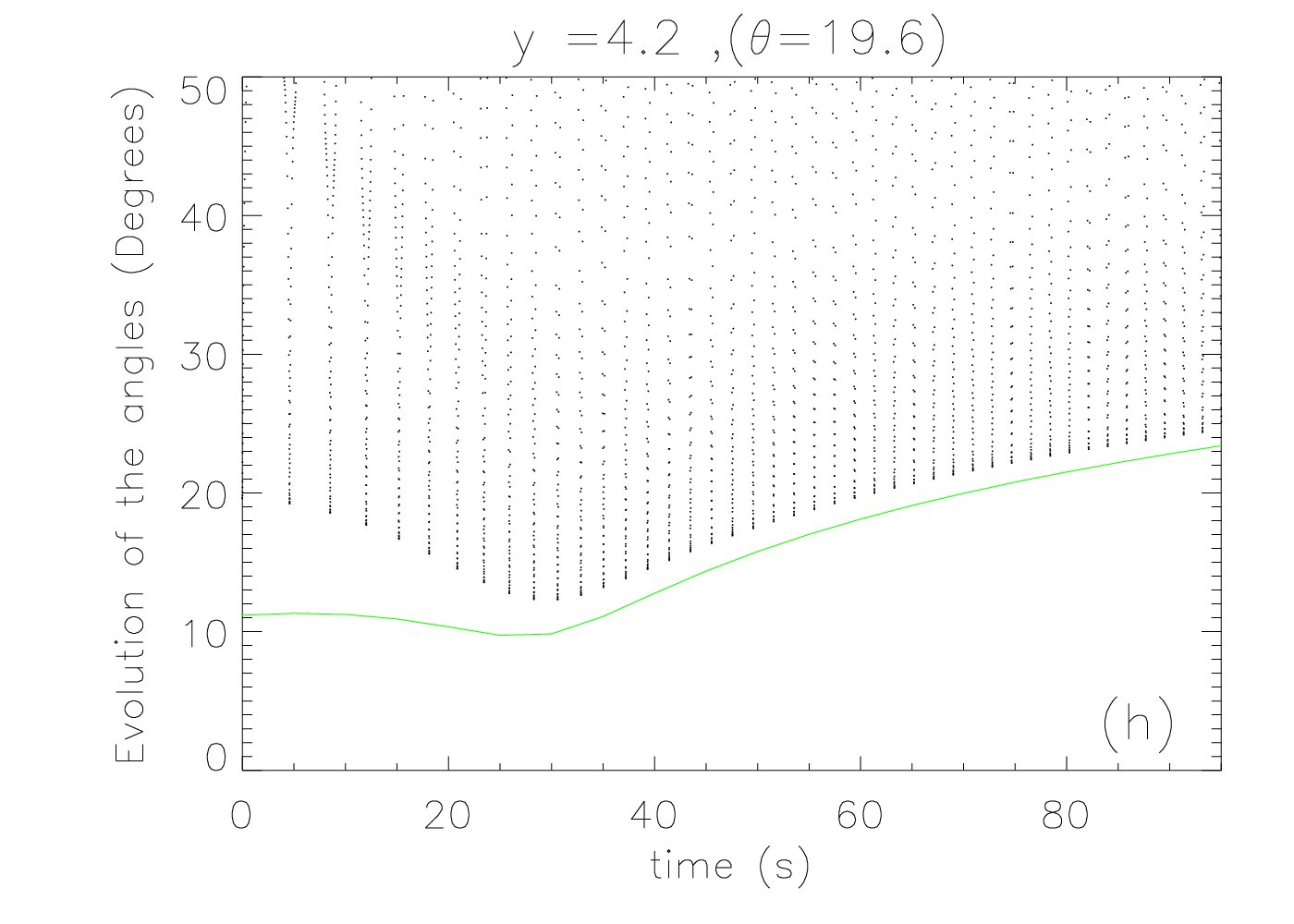}}

   \caption{The time evolution of the pitch angle $\theta$ (black dotted line) for particle orbits with initial pitch angles varying from $5^\circ$ to $19. 6^\circ$. Plot 
(a) and (b) show the pitch angle evolution for the two orbits with initial pitch angles smaller than the inital
   loss cone angle of $11.8^\circ$. Plots (c), (e), and (g) show the full time pitch angle evolution of orbits with initial pitch angles of
   $12^\circ$, $15^\circ$ and $19.6^\circ$, and (d), (f), and (h) show blow-ups of the corresponding curves. }
 \label{diffpitchangles}
    \end{figure*}

We looked next at cases of particle orbits with the same initial position and energy as orbits 1 and 2, but with initial pitch angles of
 $\theta = 5^{\circ}$, $11^{\circ}$, $12^{\circ}$, $15^{\circ}$ and $19.6^{\circ}$. This covers a range of pitch angle values starting 
inside the initial loss cone angle of $11.8^\circ$, just outside the initial loss cone angle and up to the initial pitch angle of $19.6^\circ$, which is
$180^\circ - 160.4^\circ$. The time evolution of the pitch angles for these orbits is shown in Fig. \ref{diffpitchangles}, again with
the curves for the loss cone angle (green) and the $90^\circ$ line (blue).

The orbits with initial pitch angle values $5^{\circ}$ and $11^{\circ}$ (Fig. \ref{diffpitchangles} a and b, respectively) escape from the CMT  immediately without mirroring at all.
This clearly is expected since their initial pitch angles are smaller than the initial loss cone angle.
The orbits for initial pitch angles $12^\circ$ (Fig. \ref{diffpitchangles} c), and $15^\circ$ 
(Fig \ref{diffpitchangles} e) display a number of bounces within the CMT before escaping after they cross either the lower or
the upper green curve representing the loss cone angle. Fig. \ref{diffpitchangles} (c) and (e) show the pitch angle 
evolution on the full scale, whereas Fig. \ref{diffpitchangles} (d) and (f) show blow-ups of the relevant parts of the curves close to the loss cone angle curve.
One can also see that the particle orbits display more bounces before escaping the CMT if their initial pitch angle is further away from the initial loss cone angle. 
For an initial pitch angle of $19.6^\circ$ in Fig. \ref{diffpitchangles} (g) and its blow-up (h), we see the same behaviour as already described above. 
The minima and maxima of the pitch angle curve follow the same trend as the loss cone curve, but do not cross into the loss cone, 
albeit edging slightly closer to it as time progresses.

%sav file comes from test_loss_cone_working_t_inf
%___________________________________________
%Dir:  non_rel_0_4.2_5.5_alpha_160.4_long_time
%idl: .r Loss_vs_pitch
%idl: loss_vs_pitch
%____________________________________________
    
\begin{figure*}
 \centerline{
   \includegraphics[width=6.0cm]{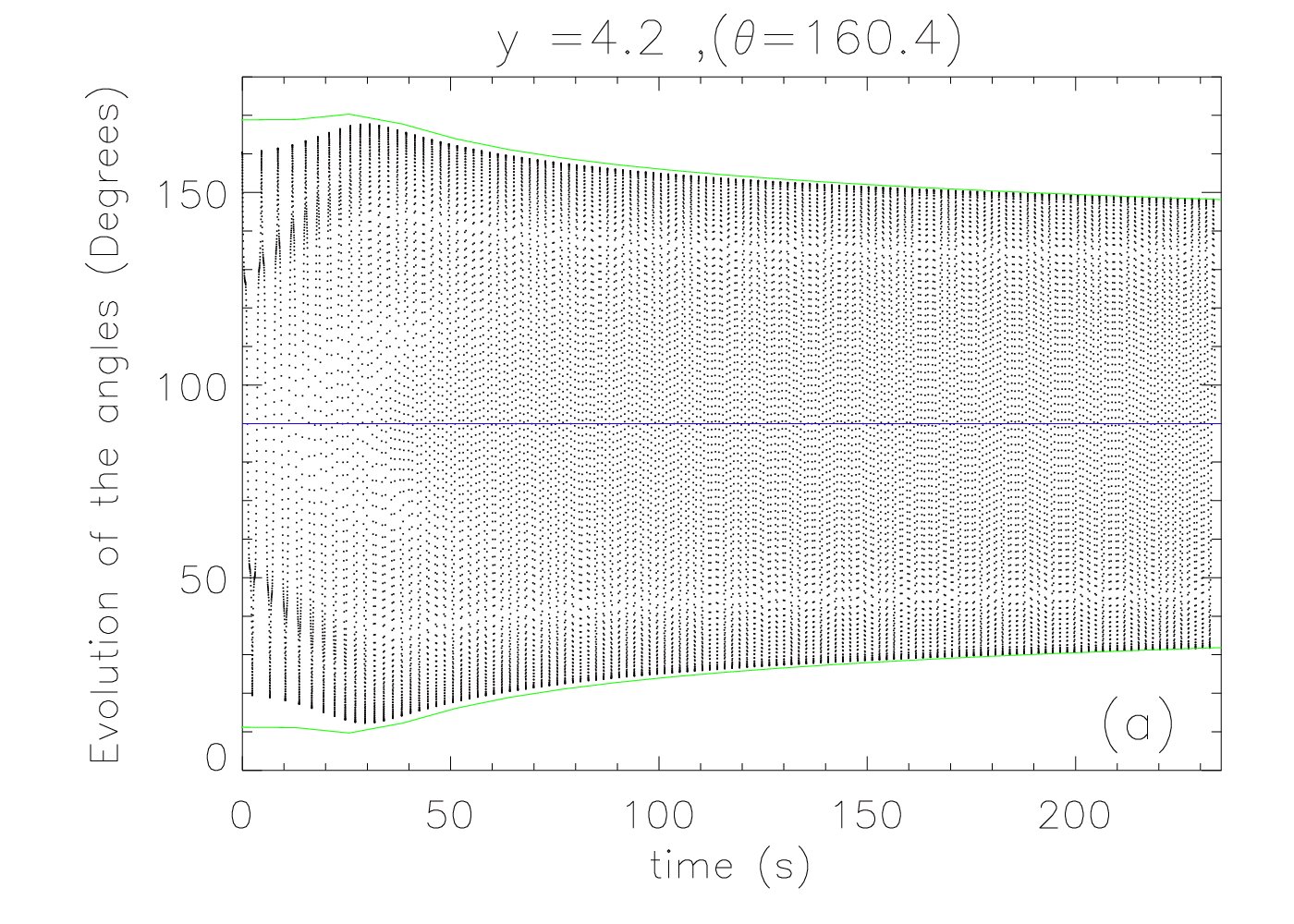}
 \includegraphics[width=6.0cm]{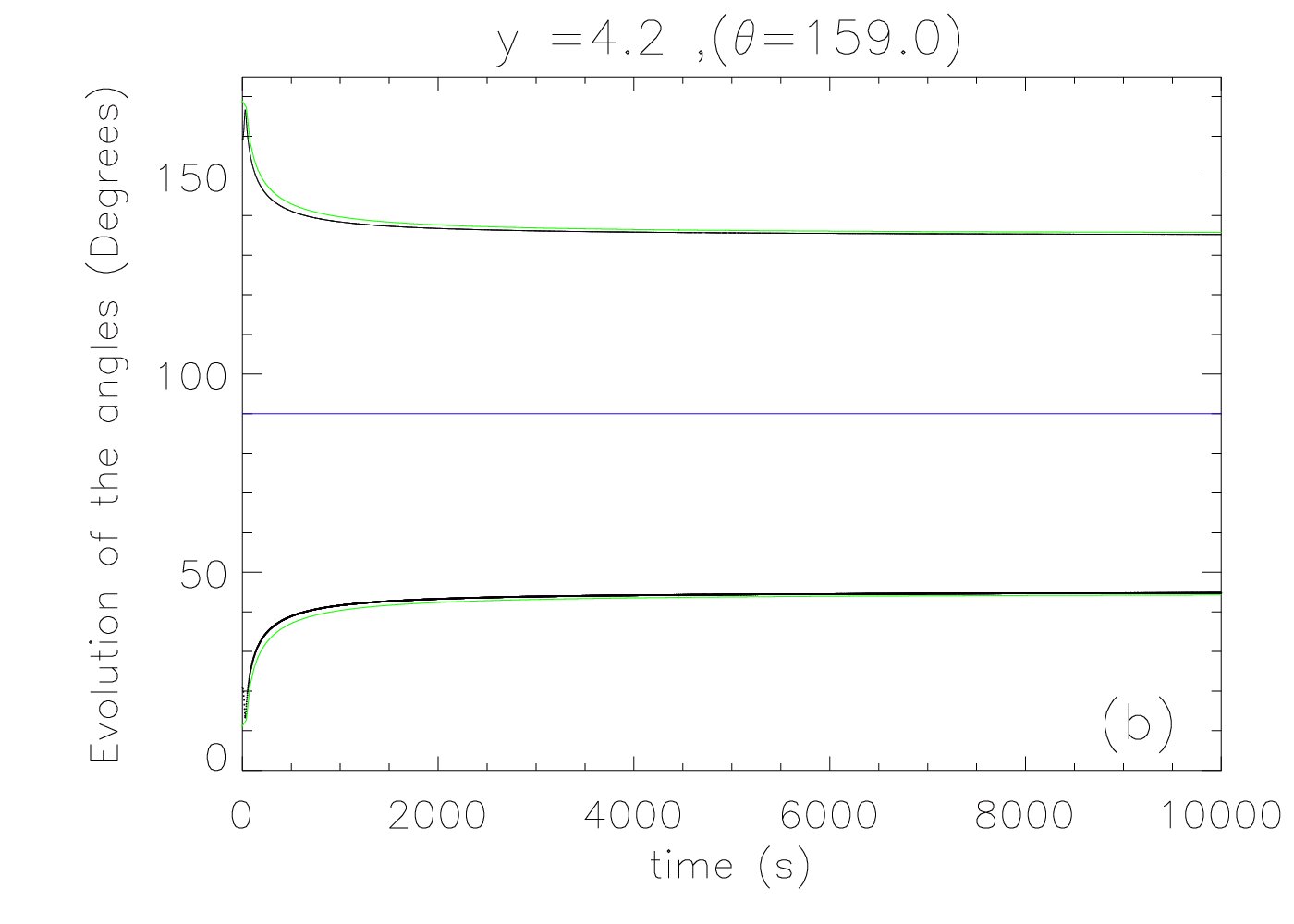}
 \includegraphics[width=6.0cm]{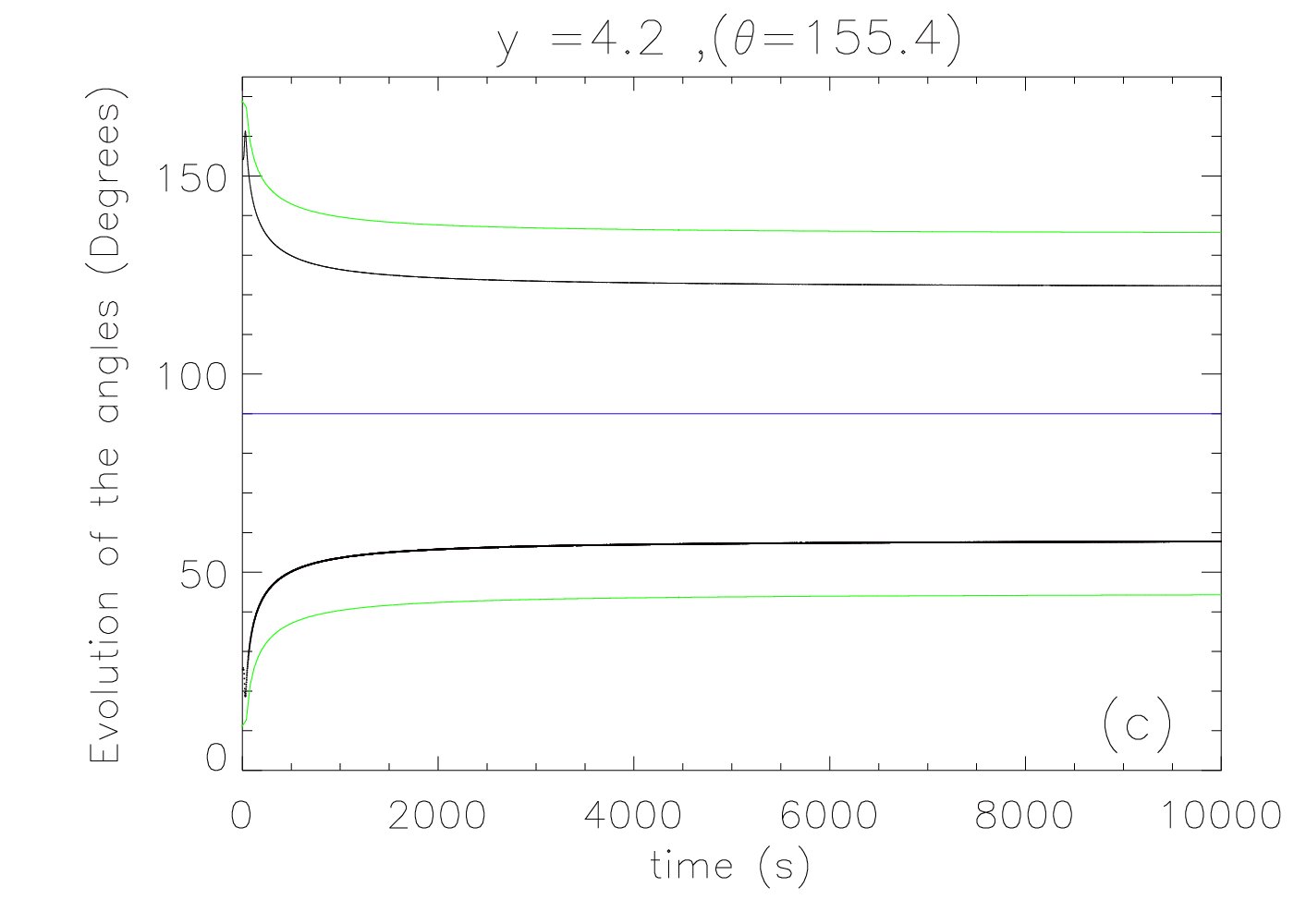}}

   \caption{{Long-term time evolution of the pitch angle for particle orbits with initial pitch angles 
    $160.4^\circ$ (subcritical, a), $159.0^{\circ}$ (critical, b), and $155.4^\circ$ (supercritical, c).
   Due to the large number of bounces, only the maxima and minima of the pitch angle curves are shown for the particle orbit plots (b) and (c). Particle orbits with subcritical angles escape, whereas particles 
  with supercritical angles and above remain trapped.}}
 \label{diffpitchangleslongtime}
    \end{figure*}

All particle orbit calculations discussed so far have been stopped at a time of $95$ seconds well before coming close to the asymptotic limit for the loss 
cone angles of  $44.8^\circ$.
This raises the question of whether the orbits trapped at $t=95$ seconds will remain trapped if the calculation is continued beyond this time. We have therefore
first increased the stopping time for the calculation of the particle orbit with an initial pitch angle of 
$160.4^\circ$ (shown in Figs. \ref{fig:particleorbits} and \ref{Figlosspitchevo} (b)) 
by a factor of approximately 100. The result of this calculation is shown in Fig. \ref{diffpitchangleslongtime} (a) and it turns out that this particle orbit eventually
escapes from the CMT at a time of about $234$ seconds (right-hand boundary of the plot).

In order to find the initial pitch angle dividing escaping particle orbits (for large $t$) from trapped particle orbits, we studied the long-term evolution of particle orbits by decreasing
the values of the initial pitch angle from $160.4^\circ$ to $155.4^\circ$. The results are shown in the other plots in Fig. \ref{diffpitchangleslongtime} (b) and (c). We would like to point out that in each of the cases, the length of the time axis is set to either the time
when the orbit escapes from the trap or to the time when the calculation is stopped ($10,000$ seconds).
For initial pitch angle values smaller than $159.0^\circ$ (Fig. \ref{diffpitchangleslongtime} (b)) we show only the maxima and minima of the pitch angle curve (the envelope of the curve created from the
values of the pitch angle when the orbits is at the apex of the field line), because of the large number of bounces the particle orbits undergo over the extended time 
period of the calculation.

As already stated above, the particle orbit with initial pitch angle of $160.4^\circ$ escapes from the CMT at around $t=234$ seconds. During further investigation a decrease in the value of the initial
pitch angle by only $0.4^\circ$ to $160.0^\circ$ already increases the time the particle orbit remains in the trap quite substantially to just over 
$400$ seconds. We also found that particle orbits with initial pitch angle below $159.0^\circ$ (Fig. \ref{diffpitchangleslongtime} (c)) seem to remain trapped, whereas orbits with initial pitch angles above $159.0^\circ$ escape.
The pitch angle evolution of the particle orbit with an initial pitch angle of $159.0^\circ$ is shown in plot (b). Within the time the calculation was run, this orbit does not escape. It thus seems that the critical initial pitch angle value dividing escaping orbits from trapped orbits for this particular combination of initial position and
energy is given by $\approx 159.0^\circ$. Particle orbits for different initial positions and energy show the same behaviour qualitatively, although there will be differences quantitatively.

 \begin{figure*}
\centerline{
     \includegraphics[width=6cm]{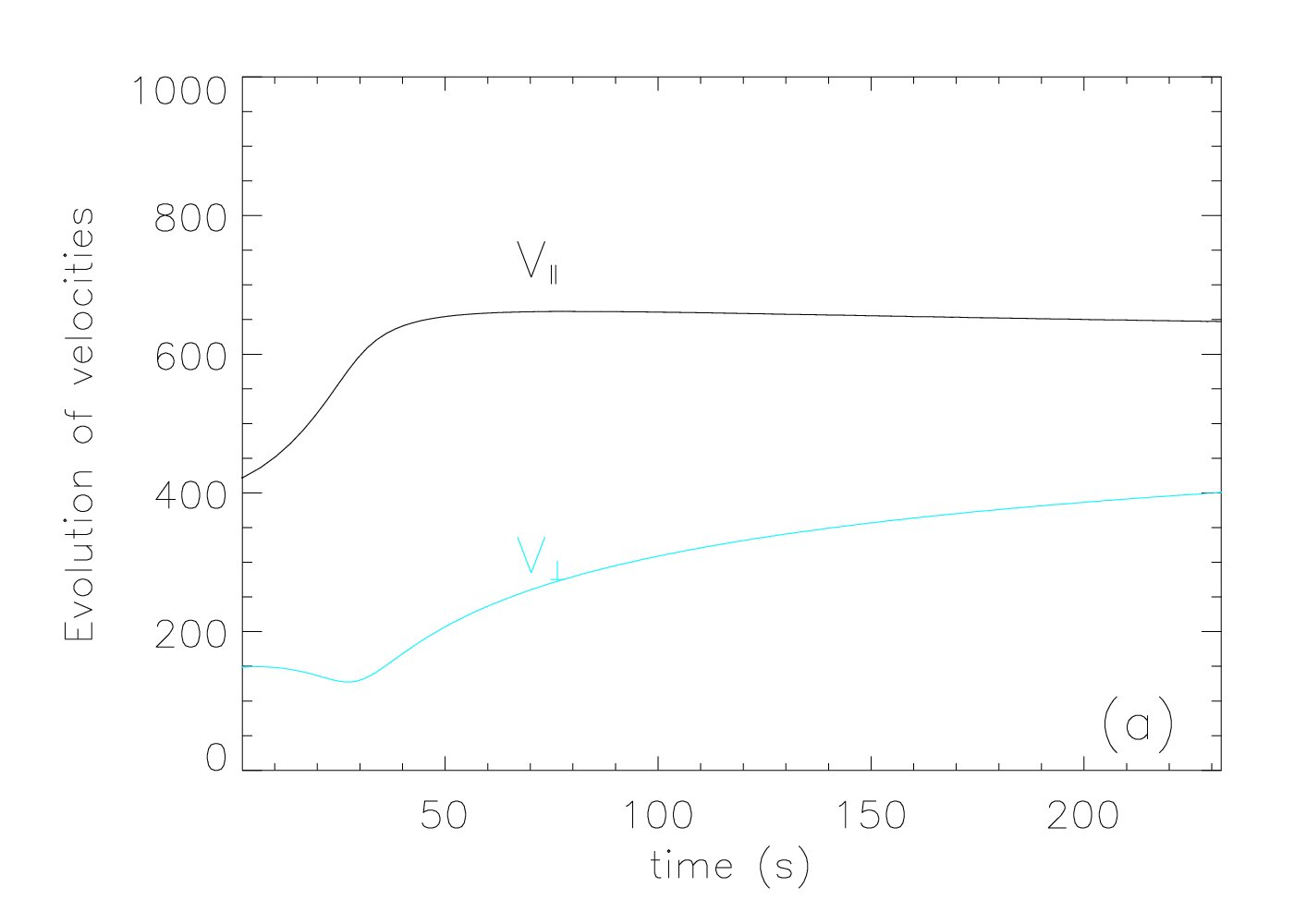}
      \includegraphics[width=6cm]{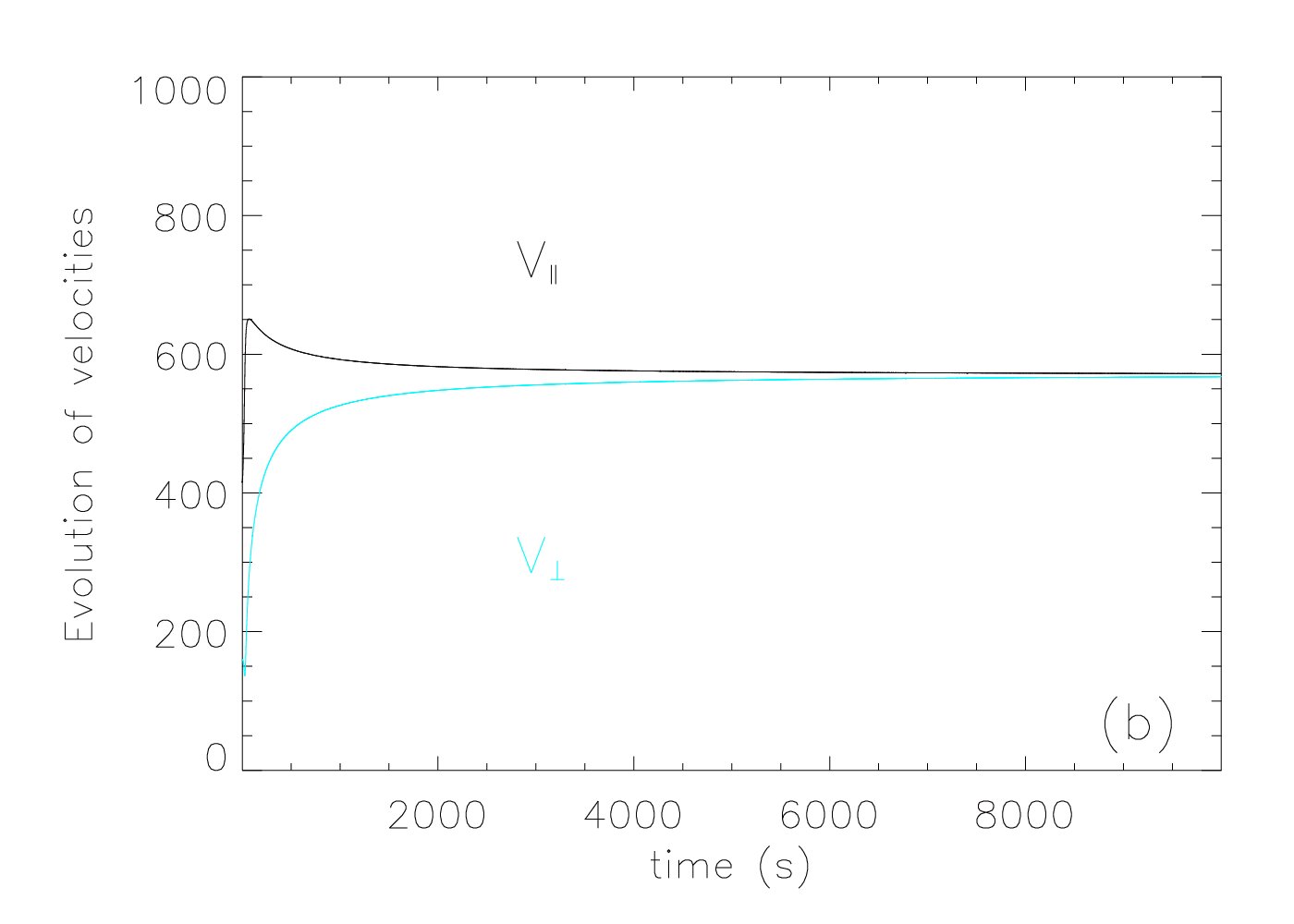}
     \includegraphics[width=6cm]{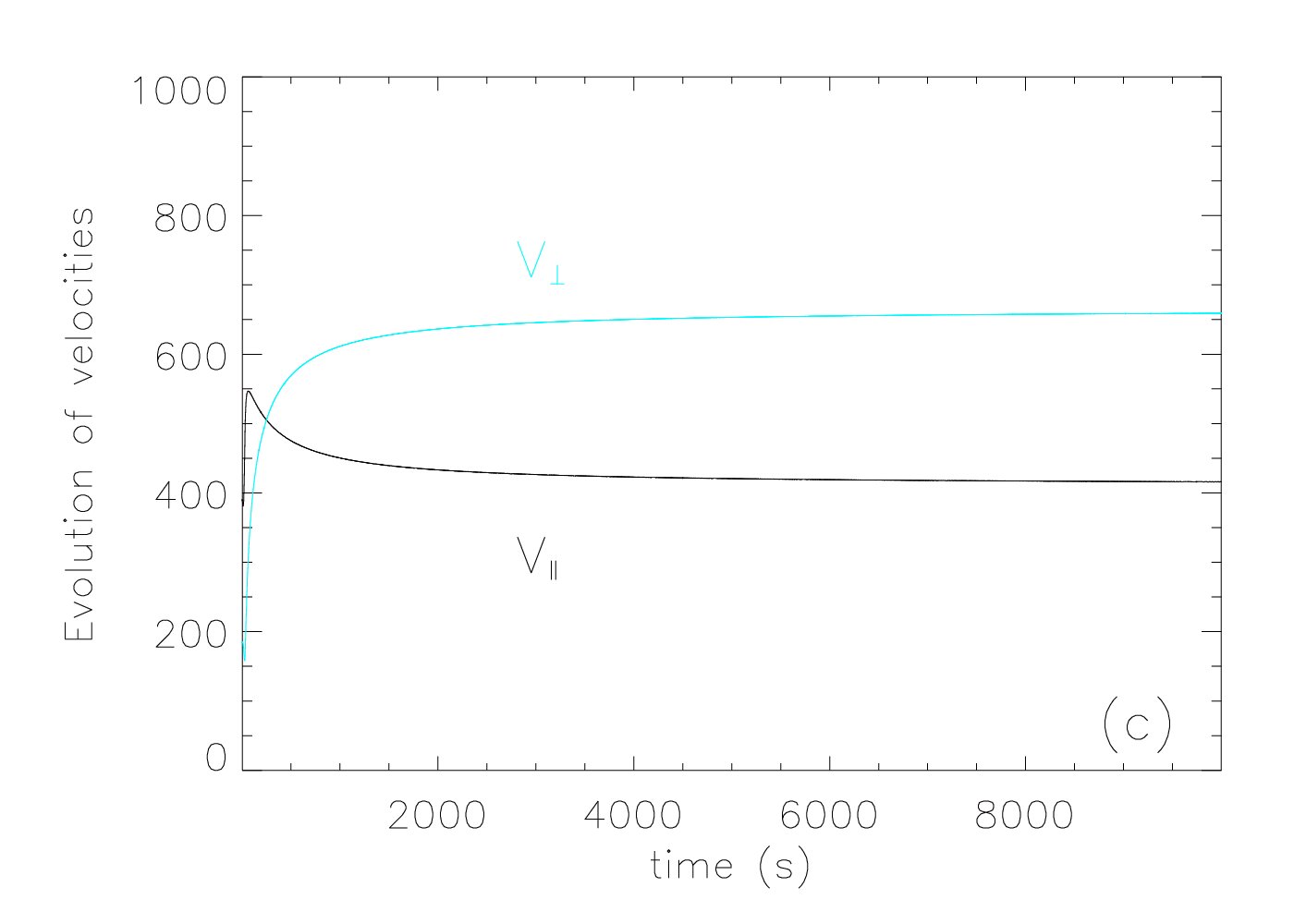}}

   \caption{{Subcritical (a), critical (b), and supercritical (c) time evolution of parallel and perpendicular velocity for the particle orbits whose pitch angle evolution 
is shown in Fig. \ref{diffpitchangleslongtime}. A key feature is that $v_\parallel$ (black curves) reaches a plateau or even decreases slightly after an initial rapid increase, while $v_\perp$ (blue) still continues to increase.}}
 \label{fig:vperp_vpar_longtime}
    \end{figure*}

Another way to look at these results is presented in Fig. \ref{fig:vperp_vpar_longtime}, which shows the time evolution of the 
parallel (black) and perpendicular velocities (blue) for the same particle orbits for which we showed the pitch angle evolution in 
Fig. \ref{diffpitchangleslongtime}. The specific feature we would like to point out in these plots is that the parallel velocity
shows a rapid increase in the initial phase of each orbit and then seems to reach a plateau or even drop off slightly, whereas the perpendicular
velocity shows a more long-term evolution although it also levels off in the long run. However, $v_\perp$ is often still increasing when
$v_\parallel$ has already either reached its plateau or is decreasing. Obviously, given the relation of the pitch angle with the components
of the particle velocity, this sheds some additional light on the situation. We hypothesise at this point that these findings could at least partially be explained
by the fact that while the motion of a field line slows down considerably relatively early in the CMT evolution and Fermi acceleration thus basically stops,
the field strength may still continue to increase due to the effect of pile up of magnetic flux from above. We will explore this hypothesis further in Sect. \ref{sec:simplemodel}
below.

\begin{figure}
\resizebox{\hsize}{!}{\includegraphics[scale=0.75]{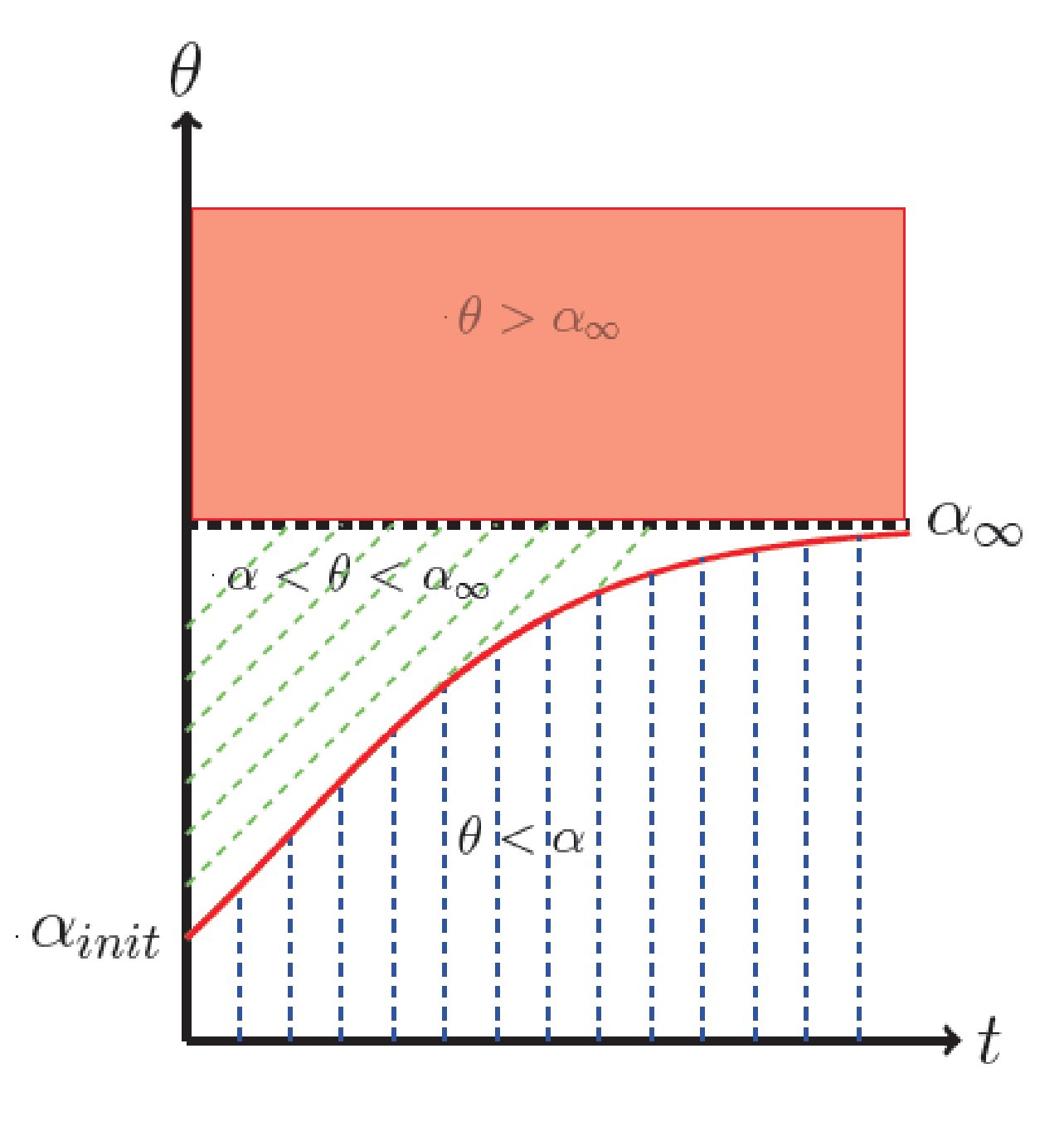}}
\caption{Sketch of the different regions in pitch angle vs time evolution in a CMT. The red curve represents the time evolution of the loss cone angle $\alpha(t)$, which
starts at a value $\alpha_{init}$ and then increases with time towards an asymptotic value $\alpha_\infty$. 
The blue hatched region represents the orbits with pitch angles less than the loss cone angle which are lost. 
Orbits with initial pitch angles greater than $\alpha_\infty$ (brown shaded region) remain trapped (although see discussion in main text). 
For orbits with initial pitch between $\alpha_{init}$ and $\alpha_\infty$ (green hatched region) the outcome depends on the relative importance of Fermi vs. Betatron
acceleration.}
\label{fig:inequalities}
\end{figure}
Generally, the situation we are facing with regards to escape or trapping of particle orbits in a CMT is summarised schematically in Fig. \ref{fig:inequalities}. The red line
indicates the time evolution of the loss cone angle $\alpha(t)$. The loss cone angle generally increases from an initial value $\alpha_{init}$ to an asymptotic value
$\alpha_\infty$ as $t\to\infty$, although as mentioned above, this increase does not necessarily have to be monotonic. We have divided the $\theta$-$t$-plane
in the sketch tentatively into three regions: a region below the red curve representing the loss cone angle $\alpha(t)$ (hatched blue), a region above the asymptotic value
of the loss cone angle, $\alpha_\infty$ (shaded brown) and the region between these two regions, where $\alpha(t) < \theta(t) < \alpha_\infty$ (hatched green).

Clearly, particle orbits whose loss cone angle crosses into the blue hatched region escape from the CMT. In particular, we found above that all orbits with initial
pitch angle below $\alpha_{init}$ are lost without even bouncing once (see Fig. \ref{diffpitchangles}). On the other hand, orbits with initial pitch angles in the brown shaded region would be expected to remain
trapped for all times, and that is consistent with our findings above. However, we have to mention that there could be a possibility of a CMT that increases the parallel energy
of particle orbits starting  with pitch angles above $\alpha_\infty$ in such a way that the pitch angle decreases with time and crosses into the blue hatched 
region. However, based on our results so far this seems to be an unlikely scenario. The most interesting region is the region hatched in green. 
As seen above, orbits starting with pitch angles in this
region can either escape from the trap if their pitch angle does not remain above the red curve during the time evolution of the orbit, 
or remain trapped indefinitely if the time evolution of
the pitch angle takes it into the brown shaded region. Our results indicate that there is a critical initial pitch angle and orbits that start with pitch angles smaller than this critical
angle will eventually cross into the blue hatched region and escape, whereas orbits with initial pitch angles above the critical angle will remain trapped. We expect that 
the behaviour we found for the specific CMT model of \citet{Giul2005} would at least qualitatively also be found for other CMT models. To corroborate this statement, we now look at how the results we found can be understood from a more basic theoretical point of view.

\section{A discussion of trapping and escape using simplified models}
\label{sec:simplemodel}

In this section, we attempt to explain the previous results regarding the trapping and escape of particle orbits in a specific CMT model using some basic 
theoretical concepts.
We will base our explanation on a simplified scenario for the acceleration within a CMT taking two features of particle orbits in CMTs found 
previously by \citet{Keith_grady_2012} into account, namely:

\begin{enumerate}
\item increases in parallel velocity occur at the apex of field lines,

\item the position mirror points moves very little during the evolution of a CMT.

\end{enumerate}

We remark that the second feature can only be satisfied approximately because otherwise no particle orbit would ever enter the loss cone. However, as we will see below,
it is an approximation that considerably simplifies the calculations without affecting the conclusions too much.

\begin{figure}

\resizebox{\hsize}{!}{\includegraphics{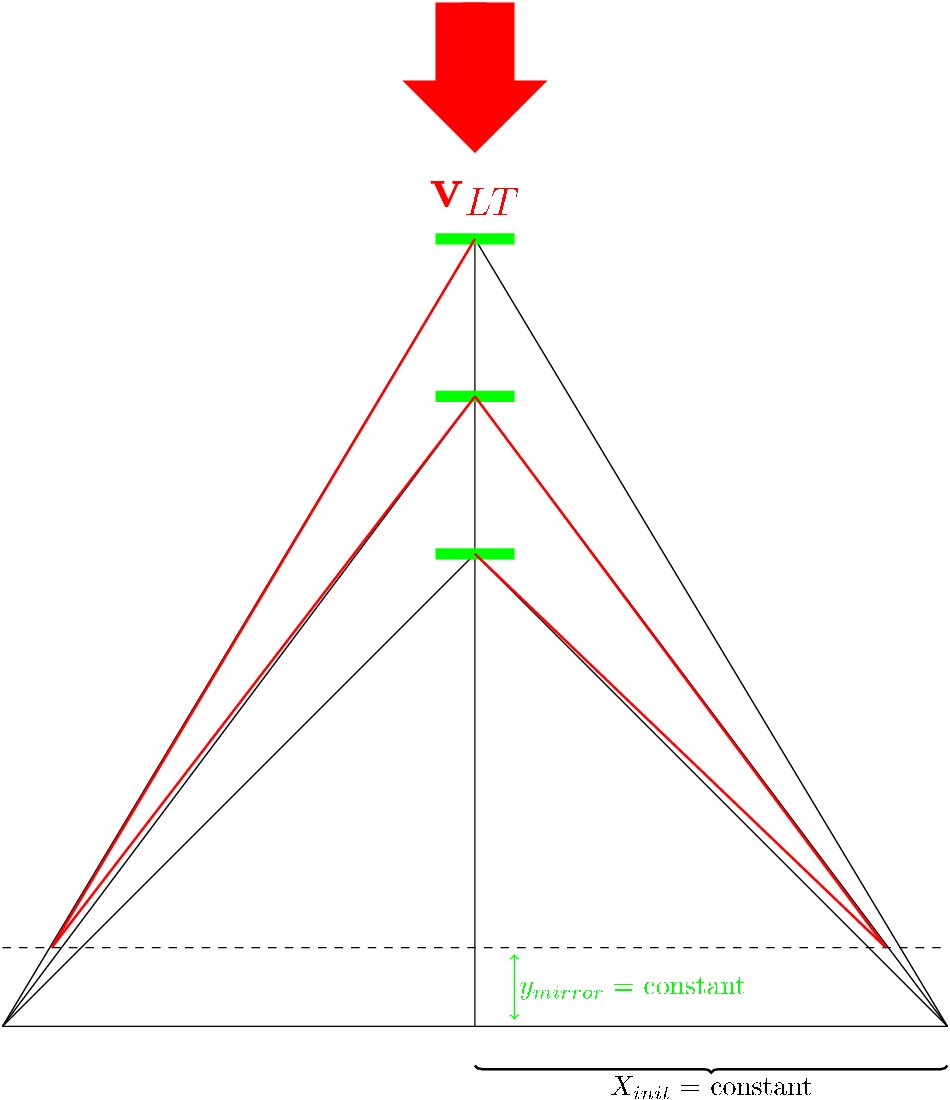}}
\caption{Sketch of the basic model used to explain the result found regarding escape and trapping of particle orbits in CMTs.}
\label{fig:simplemodel}
\end{figure}

Based on this, we simplify particle motion in a generic CMT in the way shown in the sketch in Fig. \ref{fig:simplemodel}. The particle orbit is represented by
straight lines between the mirror points, which are kept at a fixed height, and the field line apex. 
The field line apex (loop top) moves downward at a speed $v_{LT}(t)$.
The increase in parallel energy is modelled as an elastic bounce of the particle off a ``wall" moving with speed $v_{LT}(t)$ every time the particle orbit encounters the 
field line apex. We assume that the velocity remains constant between the bounces, so the simplified particle orbit consists of straight lines along which the particle moves
with constant velocity.
Simultaneously, we assume that the magnetic field strength at the field line apex evolves according to a known function $B(t)$ and that the
magnetic field strength at the footpoints, as well as their position remain constant. This allows us to
determine the time evolution of the loss cone angle for this simplified model as well as calculating the time development of the pitch angle at the field line apex just after a bounce.
 
The simplified model hence has the following ingredients:

\begin{itemize}

\item the loop top velocity $v_{LT}(t)$, which has to be specified as a function of time;

\item the loop top magnetic field strength $B(t)$, which has to be specified as a function of time;

\item the initial height of the loop top $y_{init}$;

\item the height of the mirror points $y_{mirror}$, which is assumed to be constant; and

\item the position of the foot points $x_{init}$, which is also assumed to be constant; this will be set indirectly by specifying the initial angle $\phi_{init}$
of the particle orbit with the vertical.

\end{itemize}

We also have to specify conditions for the initial velocity $v_{init}$ and the initial pitch angle $\theta_{init}$. These ingredients can be condensed into a set of simple 
algebraic equations, which we present in Appendix \ref{appendix}. These equations can be solved iteratively. 

The main task is then to make sensible choices for $v_{LT}(t)$ and $B(t)$. Obviously, guided by our CMT model discussed in Sect. \ref{sec:model}, we generally want
$v_{LT}(t)$ to be a function that decreases with time and has an asymptotic value of zero and $B(t)$ to be an increasing function of time with an asymptotic value of
$B_\infty$, for example. We remark that our CMT model, as mentioned before, has both properties for large times, but $B(t)$ is not 
necessarily monotonically increasing. We, however, have tried to keep our simplified approach as simple as possible and hence did not choose to include
this property into our analysis. To avoid making any conclusions based on just one choice of $v_{LT}(t)$ and $B(t)$, we investigated three different combinations. While we have tried to pick initial conditions and parameter values in such a way that the results resemble the numerical values of the 
particle orbits calculated kinematic
MHD CMT model  shown in Sect. \ref{sec:pitch}, one can only expect to recover the general behaviour of quantities like the pitch angle or the parallel and perpendicular
velocity components in a qualitative way and one should not expect a numerically accurate representation of the full orbit calculations.

For our first combination (model 1) we make the simplest possible choice for $v_{LT}(t)$, which is to set it equal to a constant. 
To satisfy the condition that the asymptotic value of $v_{LT}(t)$ as $t\to \infty$ should be zero, we assume that $v_{LT}(t)$ drops to zero at a 
finite time $t_{stop}$. In practice, we actually set $v_{LT}(t) =0$ when the field line apex has decreased below a fixed height $y_{min}$, 
where $y_{min} > y_{mirror}$. For this $v_{LT}(t)$, one can actually solve the algebraic equations presented in Appendix \ref{appendix} analytically.
This choice for $v_{LT}(t)$ is combined with a $B(t)$ for which we have chosen an exponentially increasing model of the form
\begin{equation}
B(t) = B_\infty - B_0 \,\exp(-t/\tau).
\label{simpleBmodel}
\end{equation}
Instead of fixing $B_\infty$ and $B_0$, we determine them from the values of $B_{start} = B(0)$ and $B_{final}= B(t_{final})$, where $t_{final}$  is the time
at which we stop the simplified model calculation.
The mirror height is determined from the initial pitch angle, according to Eq. (\ref{mirrorheight}), using $f(\theta) = \cot^{0.1} \theta$ for this and the other two
simplified models.

 \begin{figure*}
%   \begin{center}
\centerline{
   \includegraphics[width=6cm]{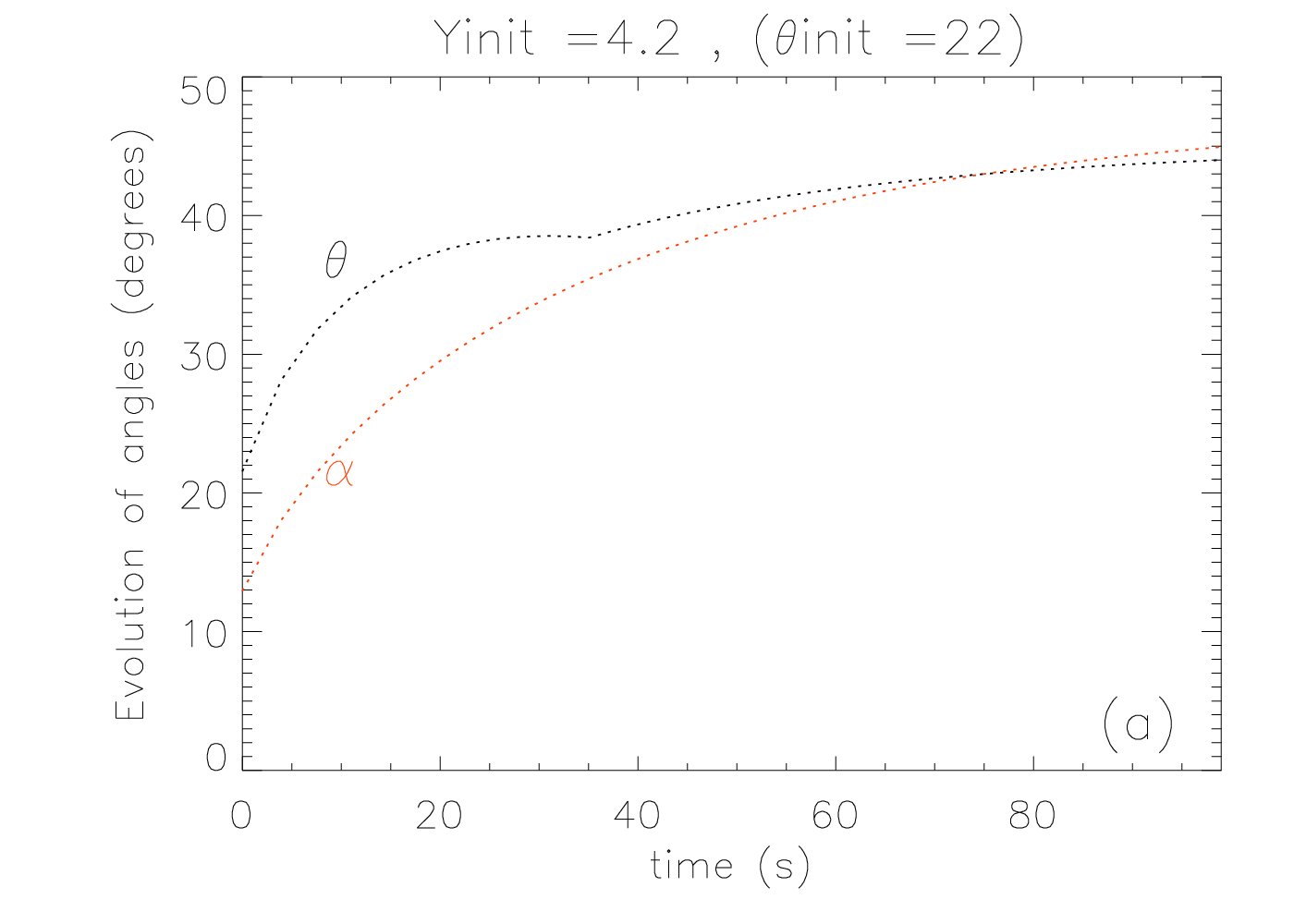}
  \includegraphics[width=6cm]{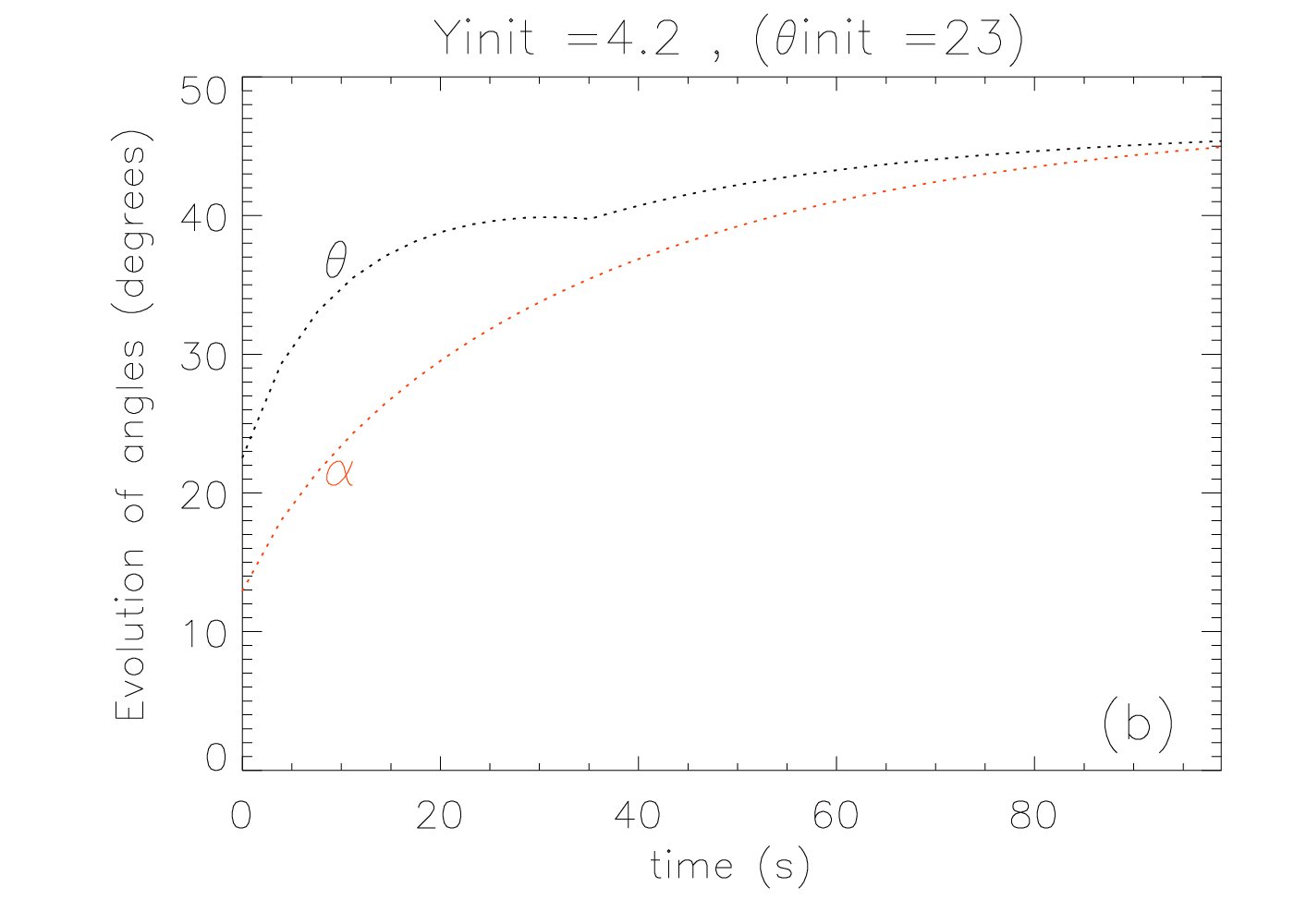}
  \includegraphics[width=6cm]{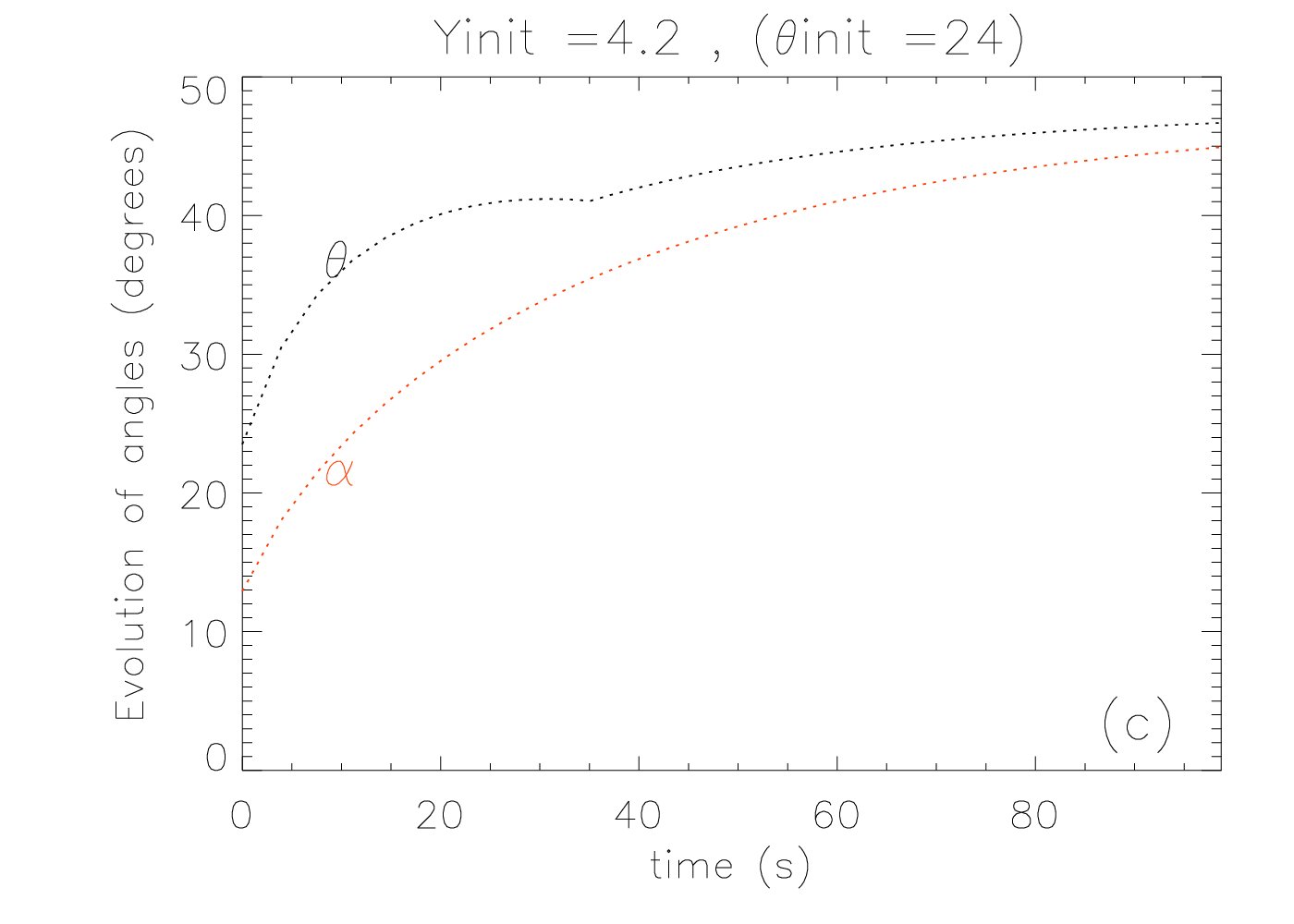}}

   \caption{{Evolution of loss cone $\alpha$ (red) and  pitch angle $\theta$ (black) for a simplified CMT acceleration model (model 1) with constant loop top velocity $v_{LT}(t)$.
Shown are the results {for} three different starting pitch angles $\theta_i$ close to the critical initial pitch angle. One can see that both 
Fermi and Betatron acceleration are operating simultaneously, however, betatron acceleration is the overall dominating mechanism, causing $\theta$ 
to increase.}}
 \label{theta_alpha_simple1_0.1}
\end{figure*}
 \begin{figure*}
%   \begin{center}
\centerline{
   \includegraphics[width=6cm]{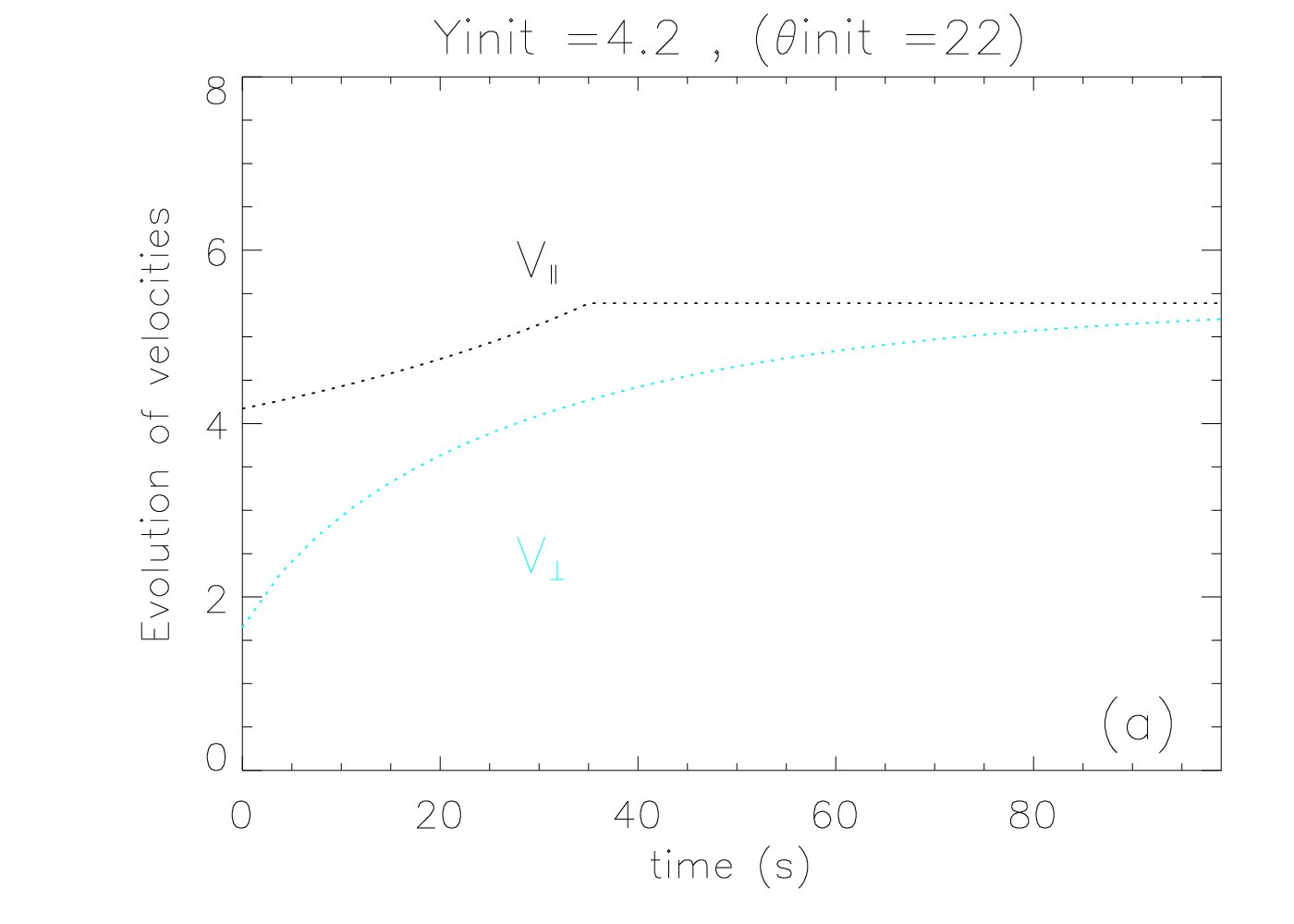}
  \includegraphics[width=6cm]{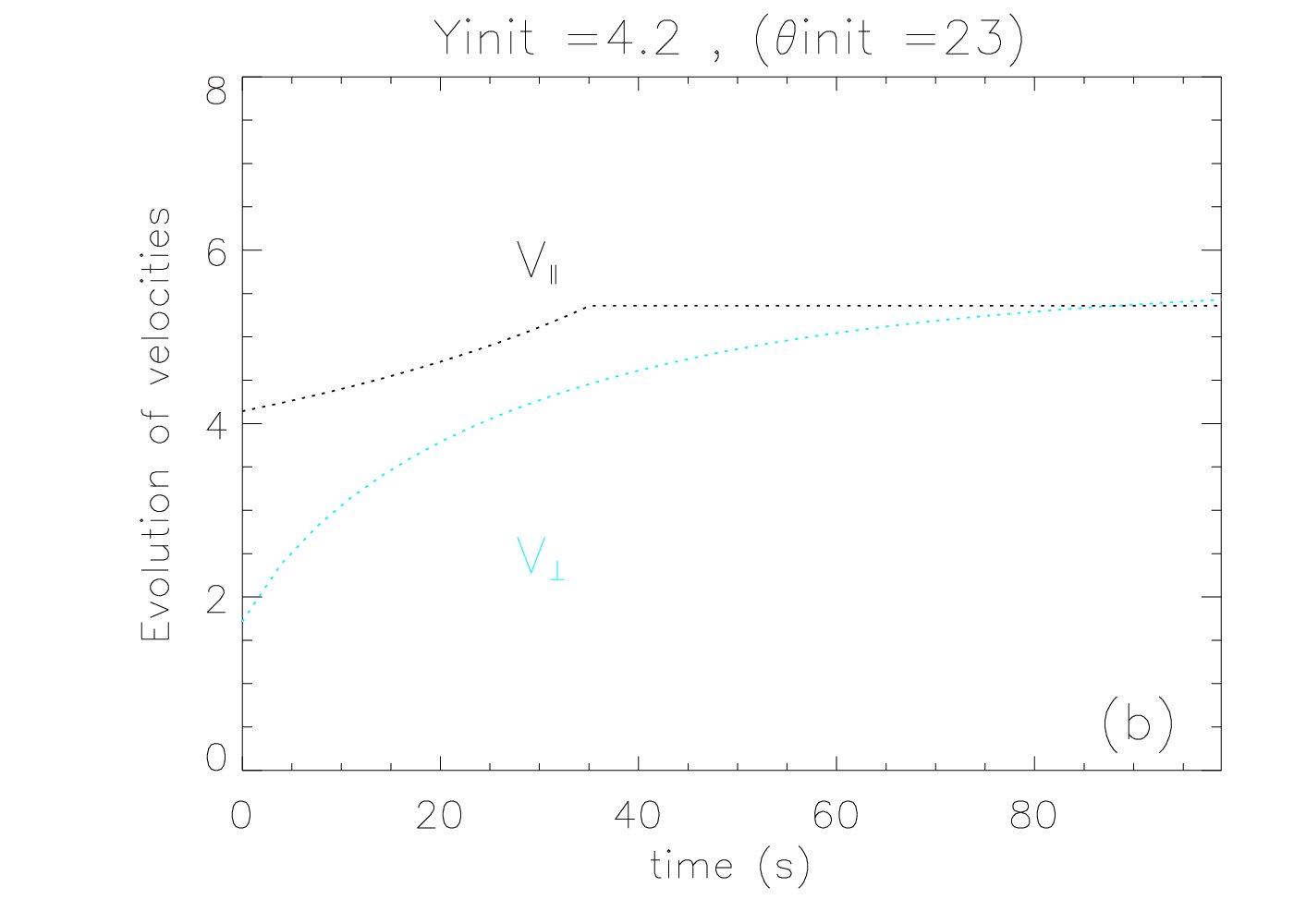}
  \includegraphics[width=6cm]{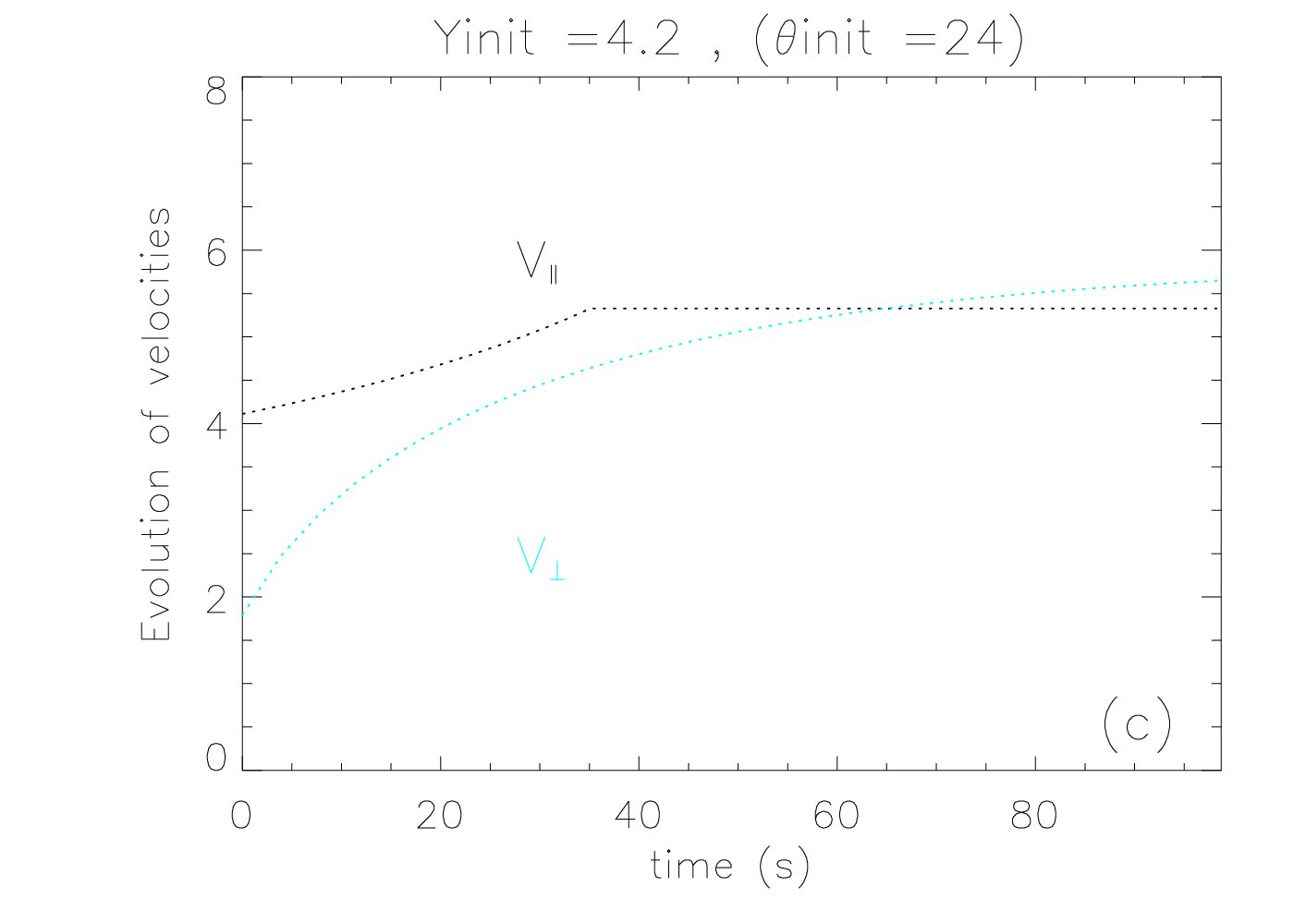}}

   \caption{{Time evolution of $v_\perp$ (blue) and  $v_\parallel$ (black) for simplified model 1. The result for three different initial pitch angles $\theta_i$ close
   to the critical initial pitch angle are shown. One can see that the parallel velocity increases sharply in the initial phase while the loop top moves and stays constant after
   the loop top stops moving.}}
   \label{vpar_vperp_simple1_0.1}
\end{figure*}

Some results for this simplified model are shown in Figs. \ref{theta_alpha_simple1_0.1} and \ref{vpar_vperp_simple1_0.1}. 
The values and parameters picked for this case were $y_{init} = 4.2 L$, $\phi_{init} = 15^\circ$, $B_{start}/B_{fp} = 0.05$, $B_{final}/B_{fp}=0.5$, $\tau = 0.4 T$, and 
$v_{LT}(t) = -4 L/T$. If we choose the same normalisation as used by \citet{Giul2005} with $L=10^7\mbox{ m}$ and $T= 100 \mbox{ s}$, we get $\tau =40 \mbox{ s}$ and 
$v_{LT}(t) = -400 \mbox{ km/s}$.
The magnetic field ratios imply that the initial loss cone angle has a value of $12.9^\circ$ and that the asymptotic loss cone angle is approximately
$45^\circ$.
The time evolution of the pitch angle at the field line apex 
compared to the loss cone angle is shown in Fig.  \ref{theta_alpha_simple1_0.1} for  three different initial pitch angles with values
of $22^\circ$, $23^\circ$, and $24^\circ$. We selected these three values, because they straddle the critical initial pitch angle, which divides trapping and escape for
this particular simplified model. In all three cases, we start with a pitch angle that is greater than the initial loss cone angle. In the case of $\theta_{init}=22^\circ$, the
curve of the pitch angle crosses into the loss cone, which means escape, whereas in the two other cases the pitch angle curves remain above the loss cone angle curve, albeit
only very slightly in the case of $\theta_{init}=23^\circ$. One can see that the pitch angle initially increases, but then reaches a maximum and 
starts to decrease (this might be difficult to see in Fig. \ref{theta_alpha_simple1_0.1}, but one can check this numerically). This is the effect of $v_{LT}(t)$ 
being non zero initially and thus Fermi acceleration becomes dominant at some point in this initial stage. After $v_{LT}(t)$ drops to zero, betatron acceleration takes over and the
pitch angle curves start increasing again.

In Fig.  \ref{vpar_vperp_simple1_0.1}, we show the time evolution of the parallel (black) and perpendicular velocities (blue) 
at the field line apex. The parallel velocity shows an initial increase during the time when $v_{LT}(t)$
is non-zero and after that it is constant. The perpendicular velocity increases on a longer timescale as it is only affected by the increase in magnetic field strength.
Despite the extreme simplification of the acceleration process in this model both Fig. \ref{theta_alpha_simple1_0.1} and Fig. \ref{vpar_vperp_simple1_0.1} show features that are qualitatively similar to the features seen in Figs. \ref{diffpitchangleslongtime} and \ref{fig:vperp_vpar_longtime}. While this is reassuring let us first see
whether this will be corroborated by the other two simplified models that we investigated.

For the second combination of $v_{LT}(t)$ and $B(t)$ (model 2), we changed $v_{LT}(t)$ to an exponentially decaying function, i.e.,
\begin{equation}
v_{LT}(t) = v_0 \exp(-t/\tau_{v}),
\label{vltexpmodel}
\end{equation}
with $v_0$ being the initial loop top velocity and $\tau_{v}$ the timescale on which $v_{LT}(t)$ drops off. This timescale is not necessarily the same as the
timescale $\tau$ on which $B(t)$ increases (see Eq. (\ref{simpleBmodel})), and we have chosen $\tau_v = 0.144T$ and $\tau=0.25T$ (corresponding
to $14.4$ seconds and $25$ seconds in the \citet{Giul2005} normalisation). 
All initial conditions and the $B$-field ratios as well as $\phi_{init}$ have been chosen to remain the same as for the first case.
Because the loop top velocity is now exponentially decreasing, we have chosen its initial velocity to be higher than the constant velocity in model 1 with $v_0=-10$ 
($=-1000 \mbox{ km/s}$ for the initial velocity).
 
\begin{figure*}

\centerline{
   \includegraphics[width=6cm]{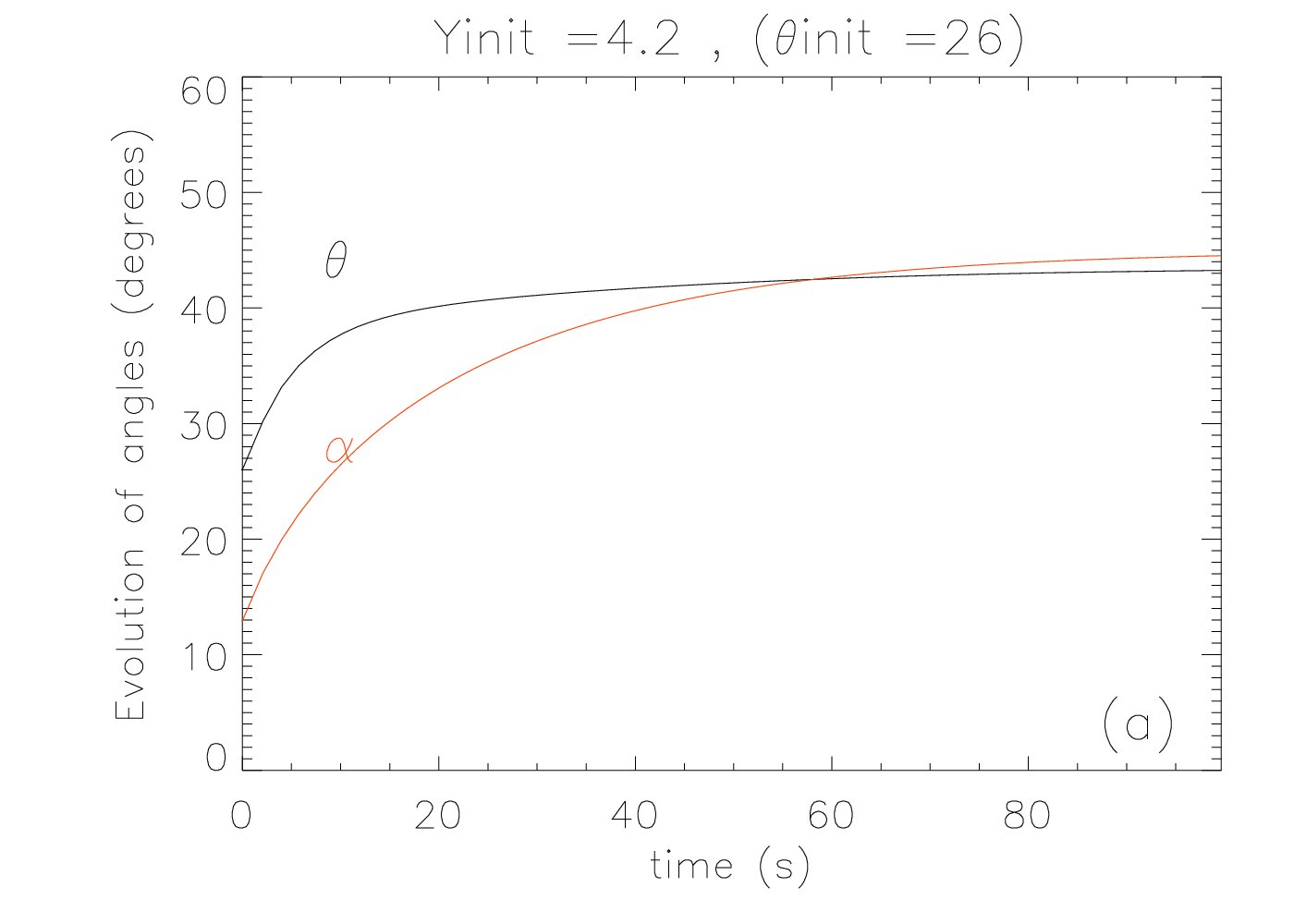}
  \includegraphics[width=6cm]{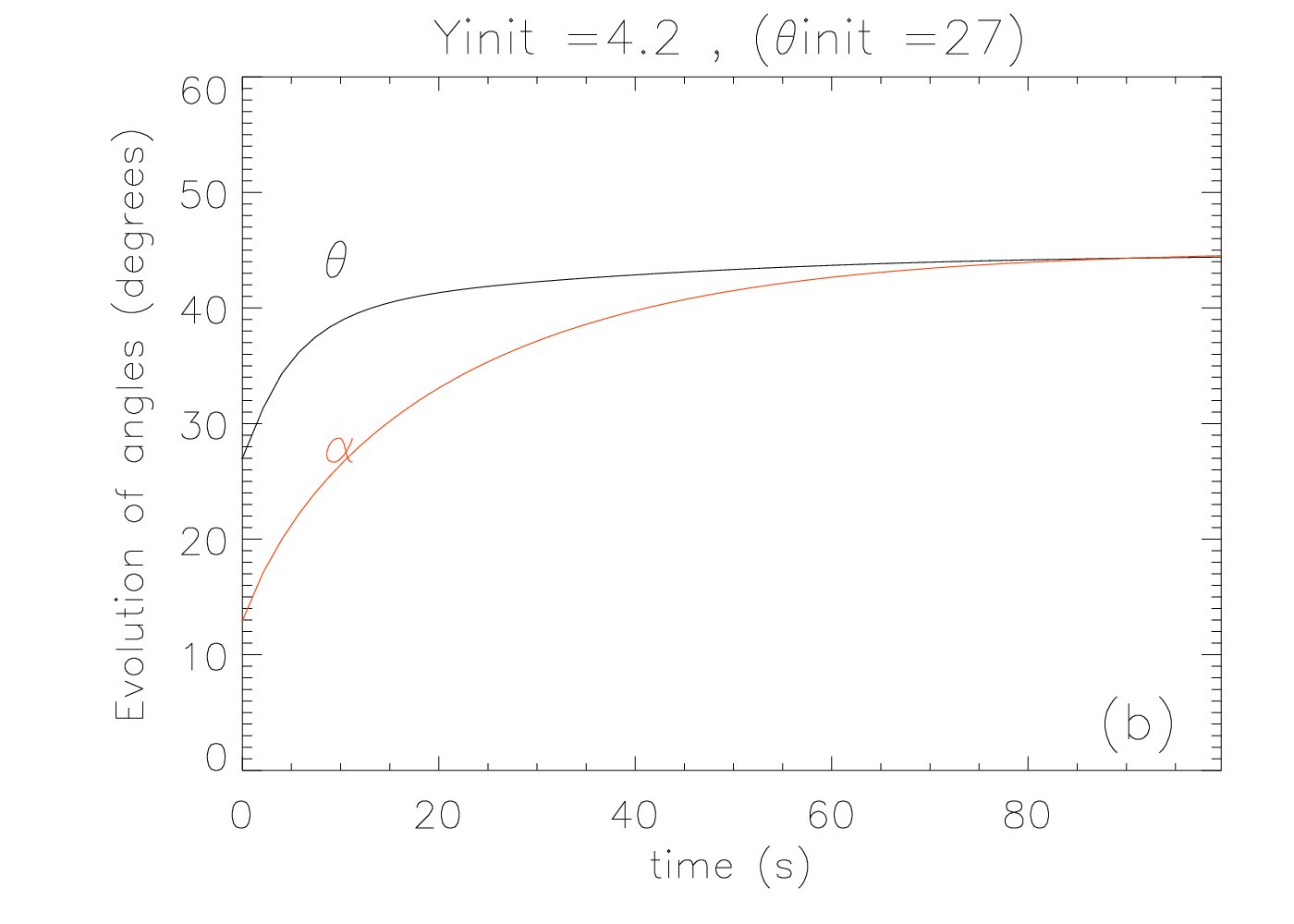}
  \includegraphics[width=6cm]{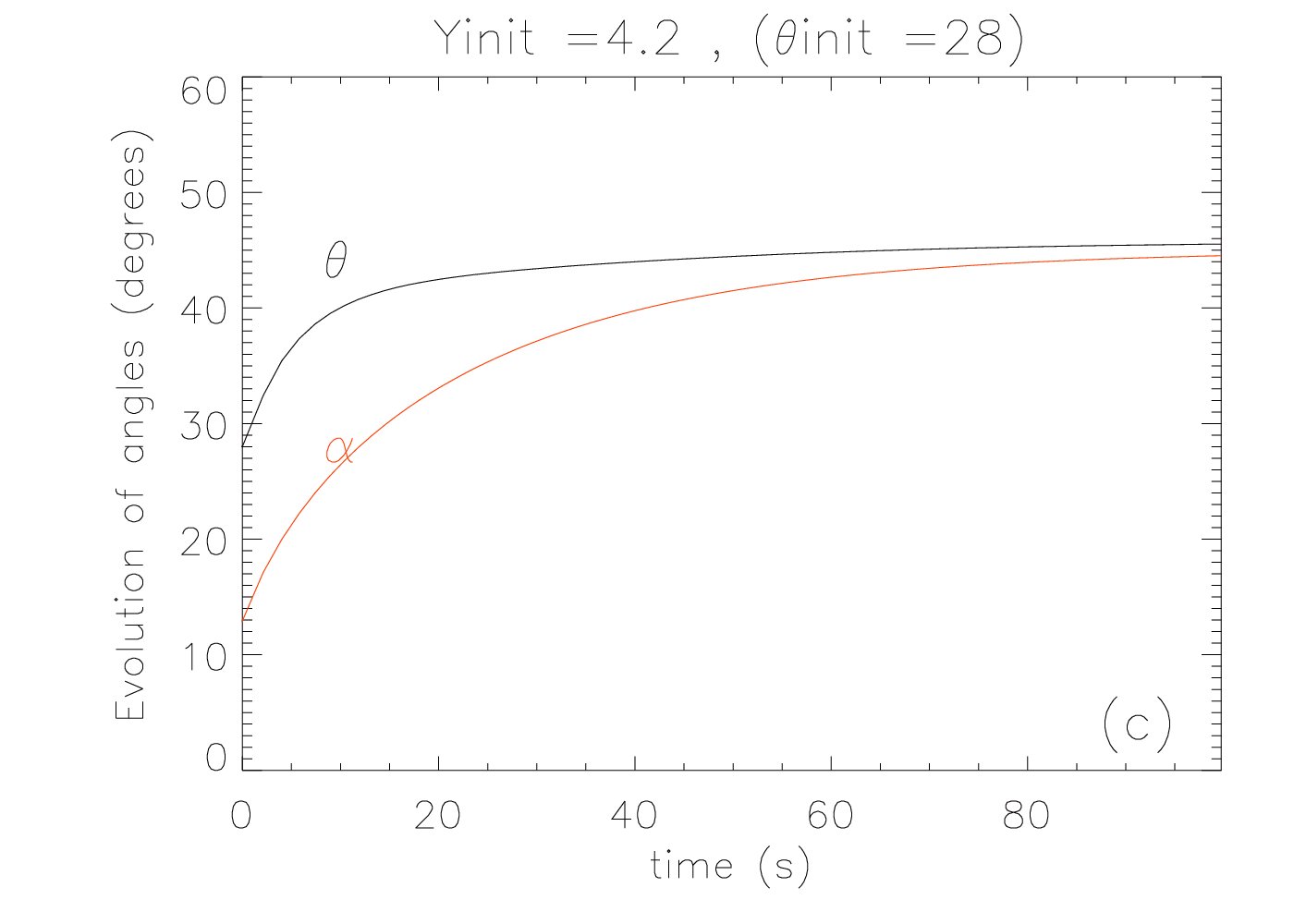}}

\caption{{Same quantities as in Fig. \ref{theta_alpha_simple1_0.1}, but for model 2. For a detailed discussion, 
see main text.}}
 \label{theta_alpha_simple2_0.1}
\end{figure*}
 \begin{figure*}
%   \begin{center}
\centerline{
   \includegraphics[width=6cm]{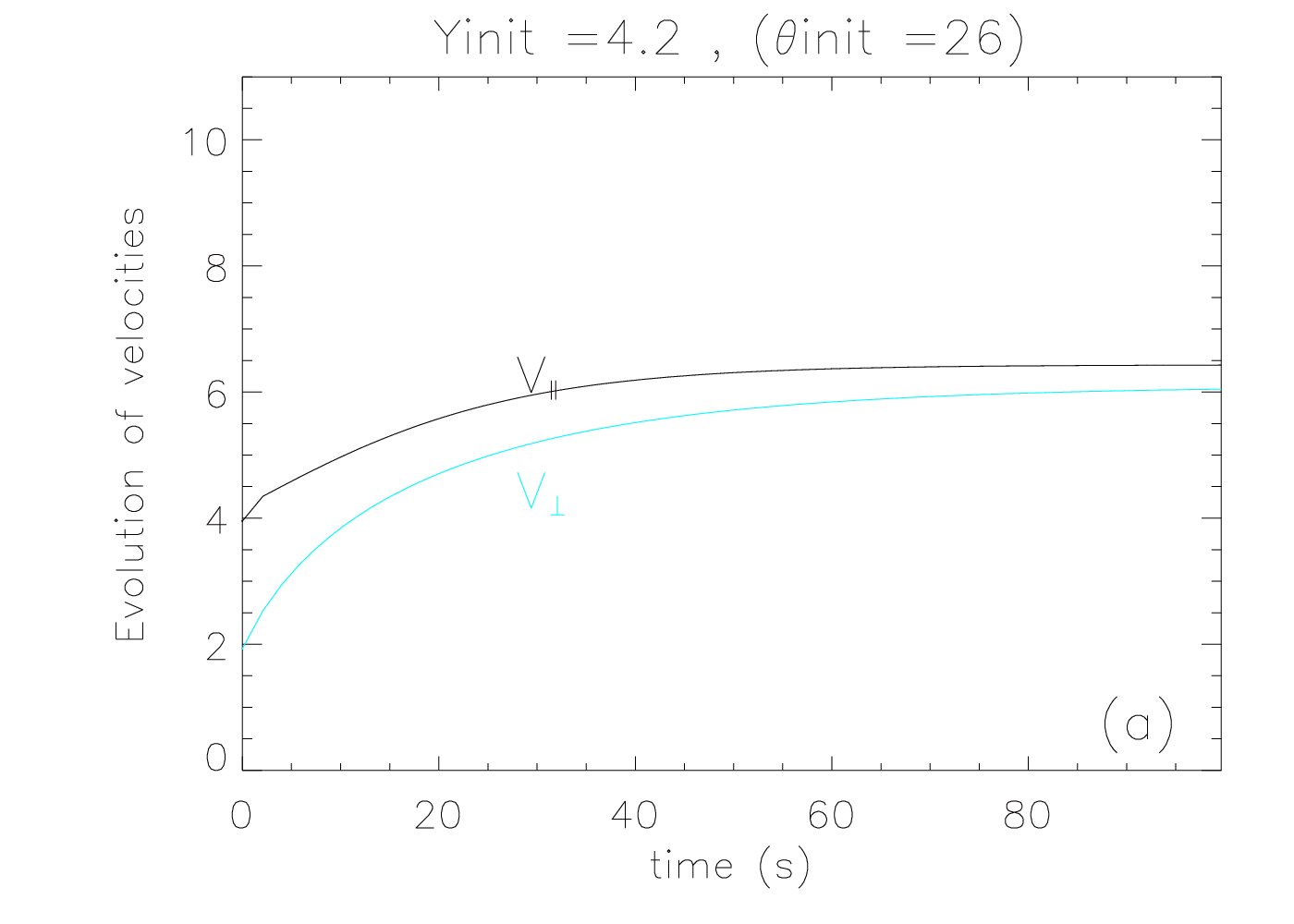}
  \includegraphics[width=6cm]{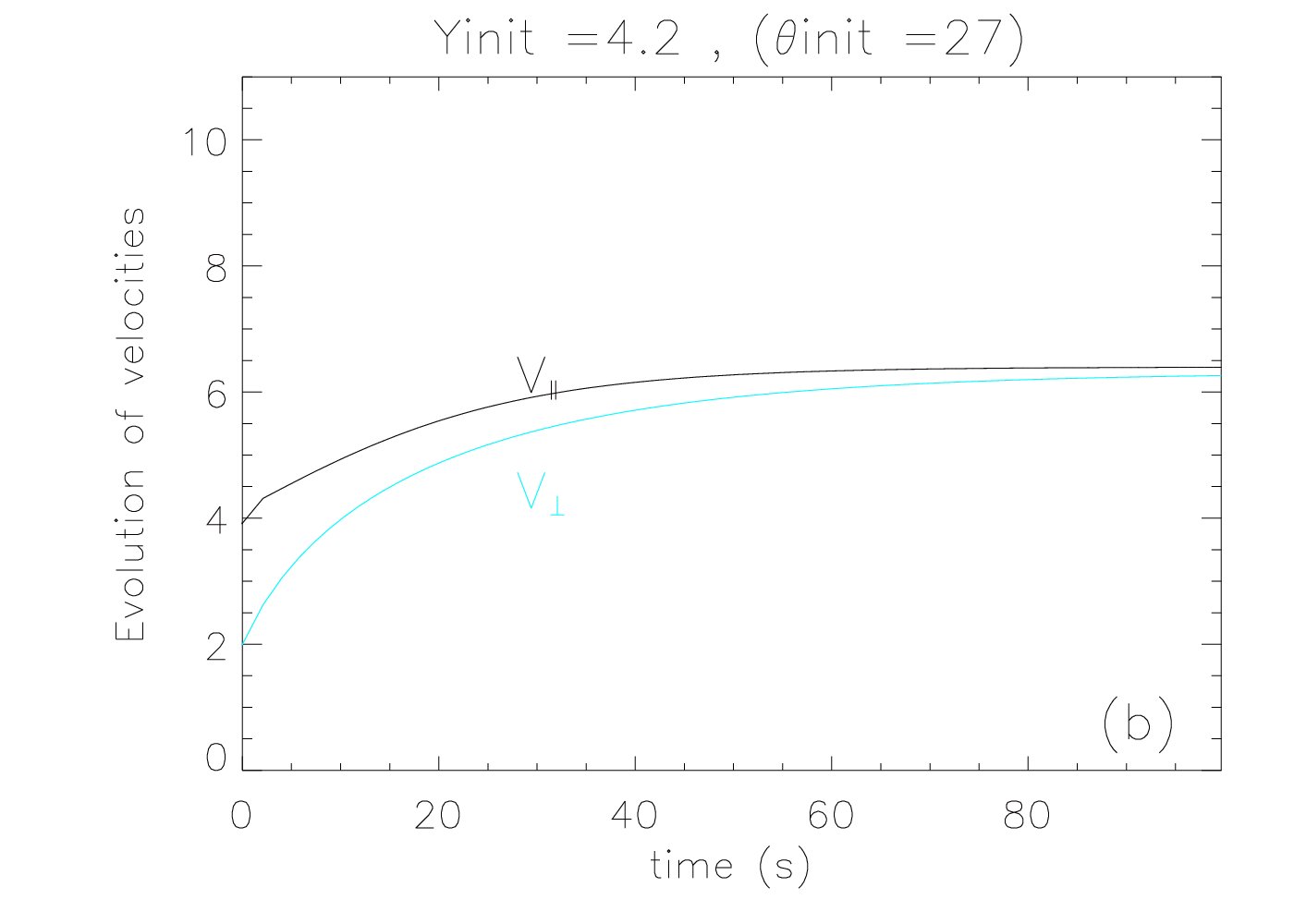}
  \includegraphics[width=6cm]{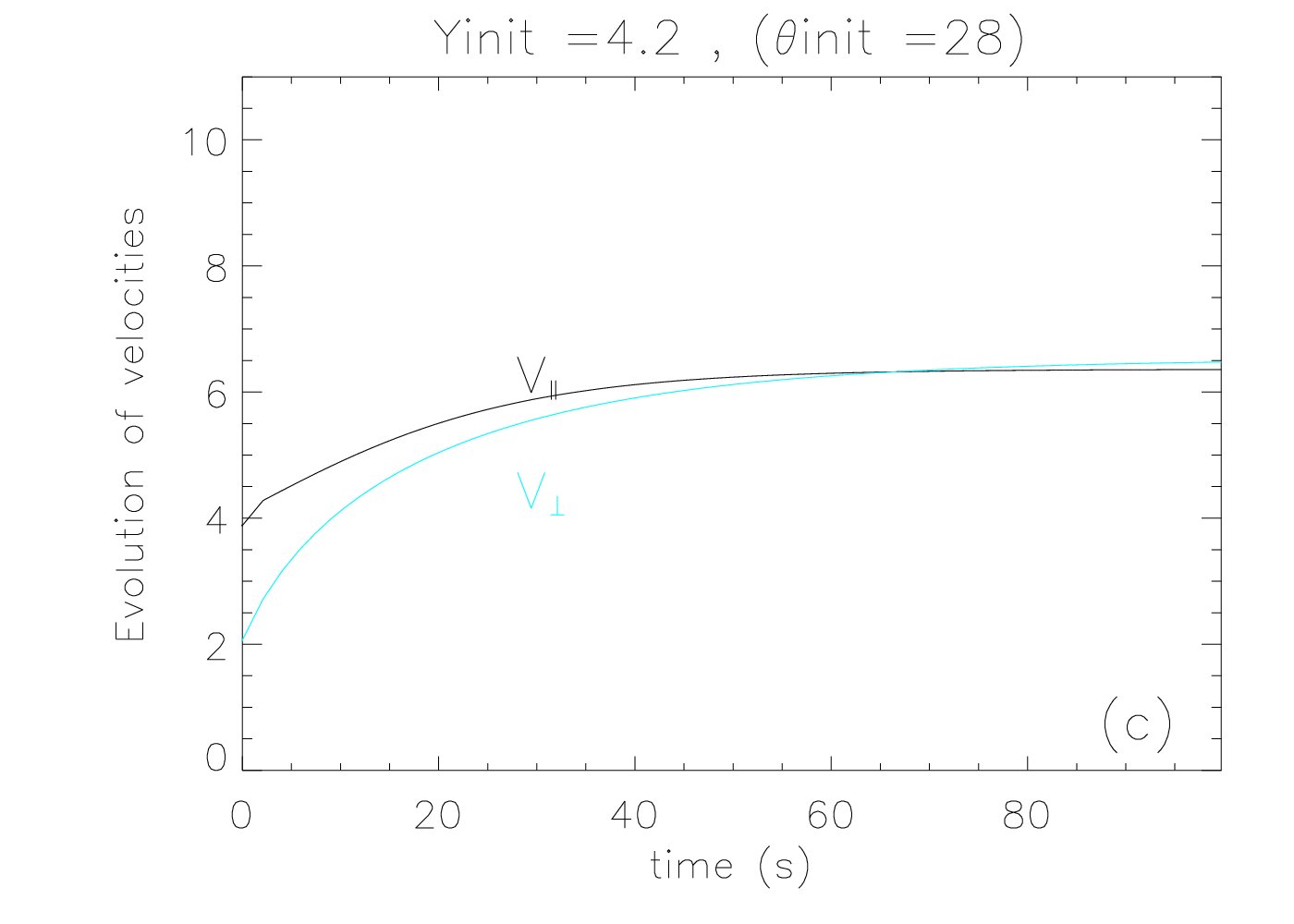}}

   \caption{{Same quantities as in Fig. \ref{vpar_vperp_simple1_0.1}, but for model 2. For a detailed discussion, see main text.}}
      \label{vpar_vperp_simple2_0.1}
\end{figure*}

The results for model 2 are shown on Figs. \ref{theta_alpha_simple2_0.1} and \ref{vpar_vperp_simple2_0.1}. We again show plots for cases
with initial pitch angles close to the critical pitch angle defining the boundary between escape and trapping ($\theta_{init} = 26^\circ$, $27^\circ$, and $28^\circ$).
The values of these initial pitch angles obviously differ from the values in model 1, because the time evolution of case 2 differs. Qualitatively, however, we again see
similar features as in the full orbit calculations, with e.g., $v_\parallel$, showing a stronger increase at the beginning when $v_{LT}(t)$ is large and then levelling off.

For the third case of a simplified model (model 3), we choose to take the functions given by \citet{Asch_2004} for $v_{LT}(t)$ and $B(t)$, but modify them slightly 
so that they match our desired model features. In particular, we assume that the asymptotic $B(t)$ as $t\to\infty$ will be smaller than the footpoint field
strength, whereas  \citet{Asch_2004}  assumes that they are the same, implying that the asymptotic loss cone angle approaches $90^\circ$.
We also start with a finite magnetic field strength, whereas Aschwanden's model starts with vanishing field strength. In practice we achieve that
by simply starting the model equations at a finite time of $t >0$.
The loop top velocity function $v_{LT}(t)$ actually has the same form as Eq. (\ref{vltexpmodel}) and thus does not have to be changed. The loop top
magnetic field strength is given by
\begin{equation}
B(t) = (B_\infty -B_0)\sin^2\left[  \frac{\pi}{2} \left( 1 - \exp(-t/\tau) \right)\right] + B_0,
\end{equation}
so that the magnetic field strength at $t=0$ is $B_0$ and as $t\to\infty$ we find $B(t) \to B_\infty$. We choose the same magnetic field ratios as for the two previous cases
and also keep $\tau_v$ and $\tau$ the same as in case 2. Similarly, all other parameters and initial conditions remain unchanged.

\begin{figure*}
%   \begin{center}
\centering
   \includegraphics[width=6cm]{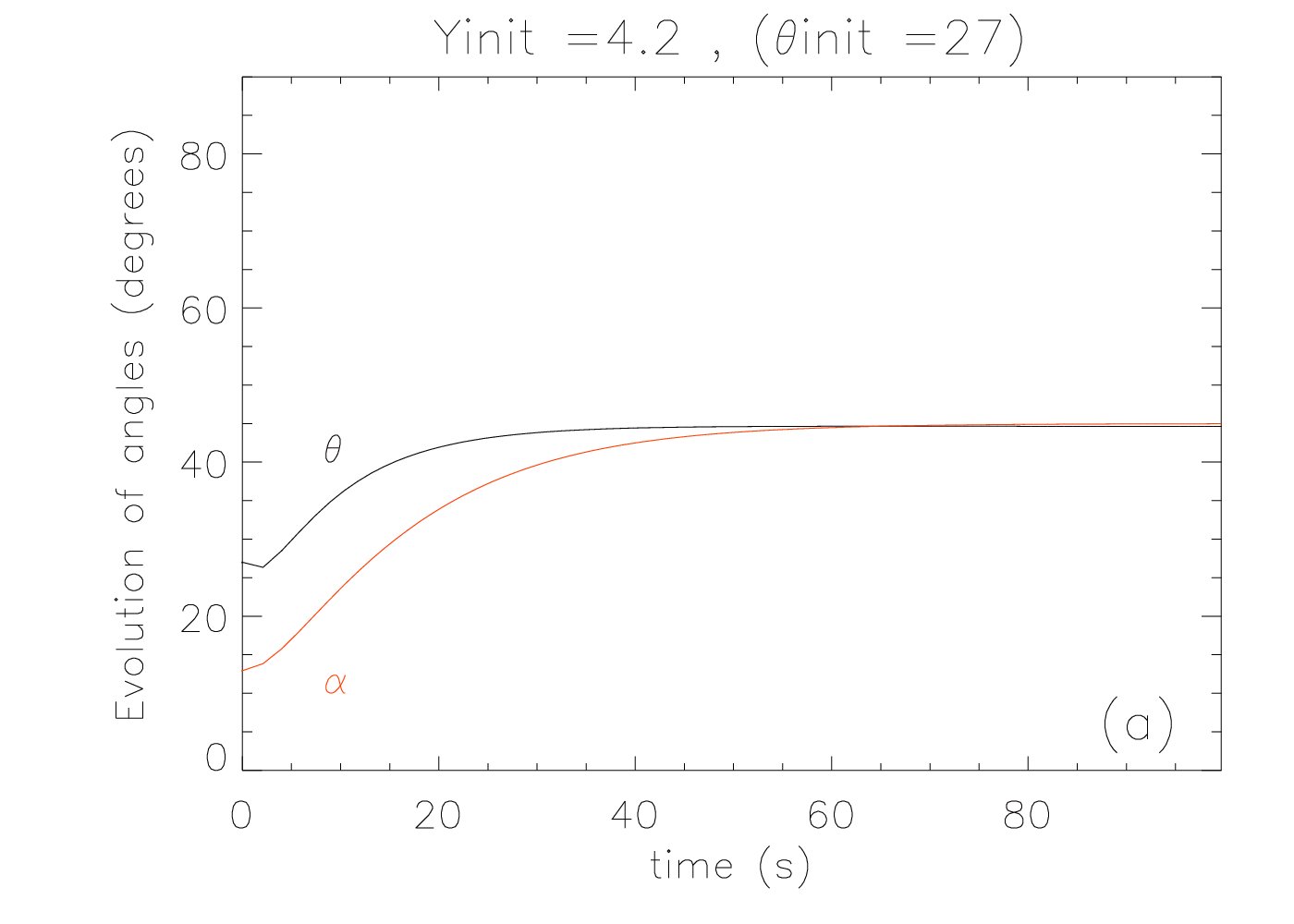}
  \includegraphics[width=6cm]{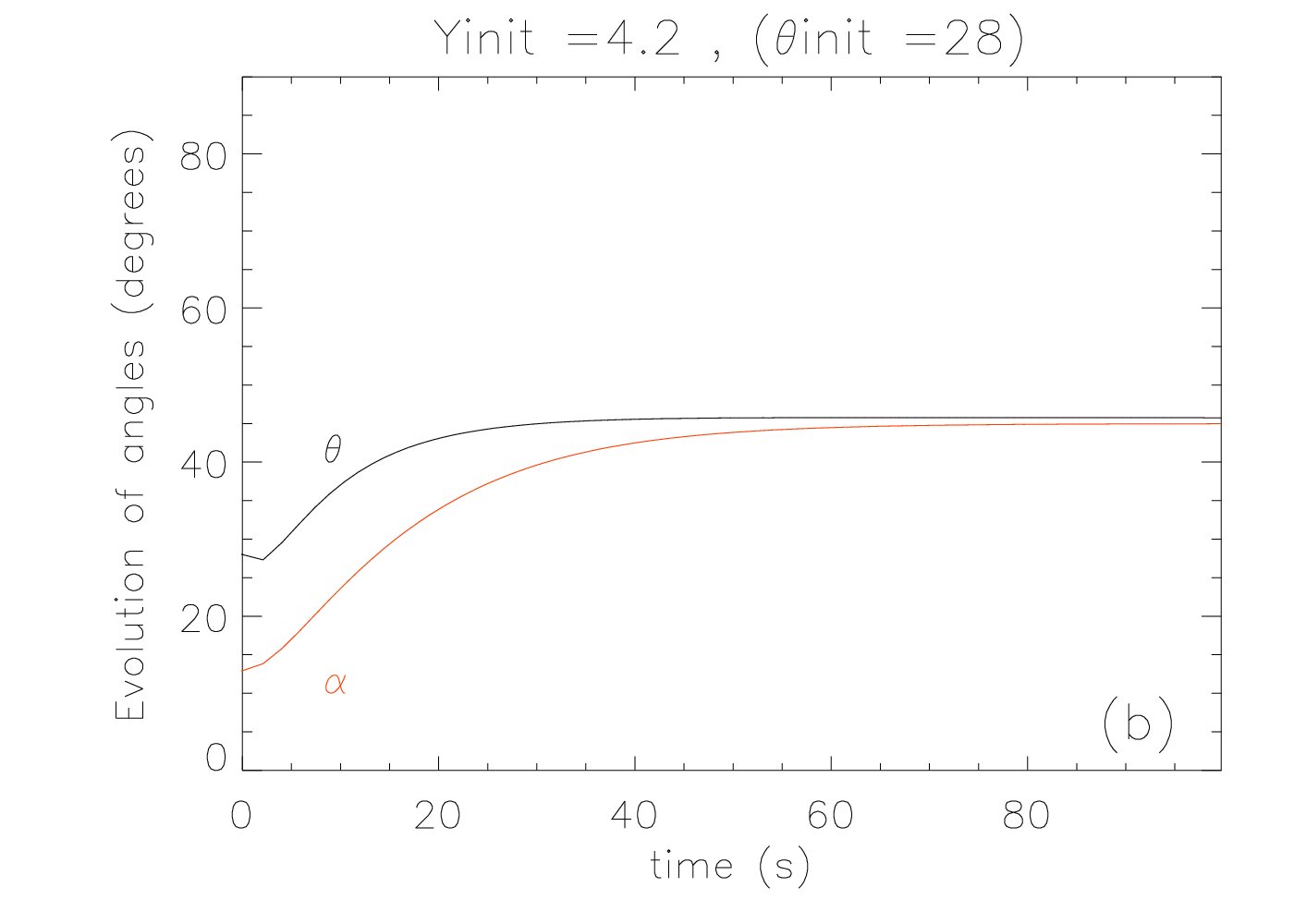}
  \includegraphics[width=6cm]{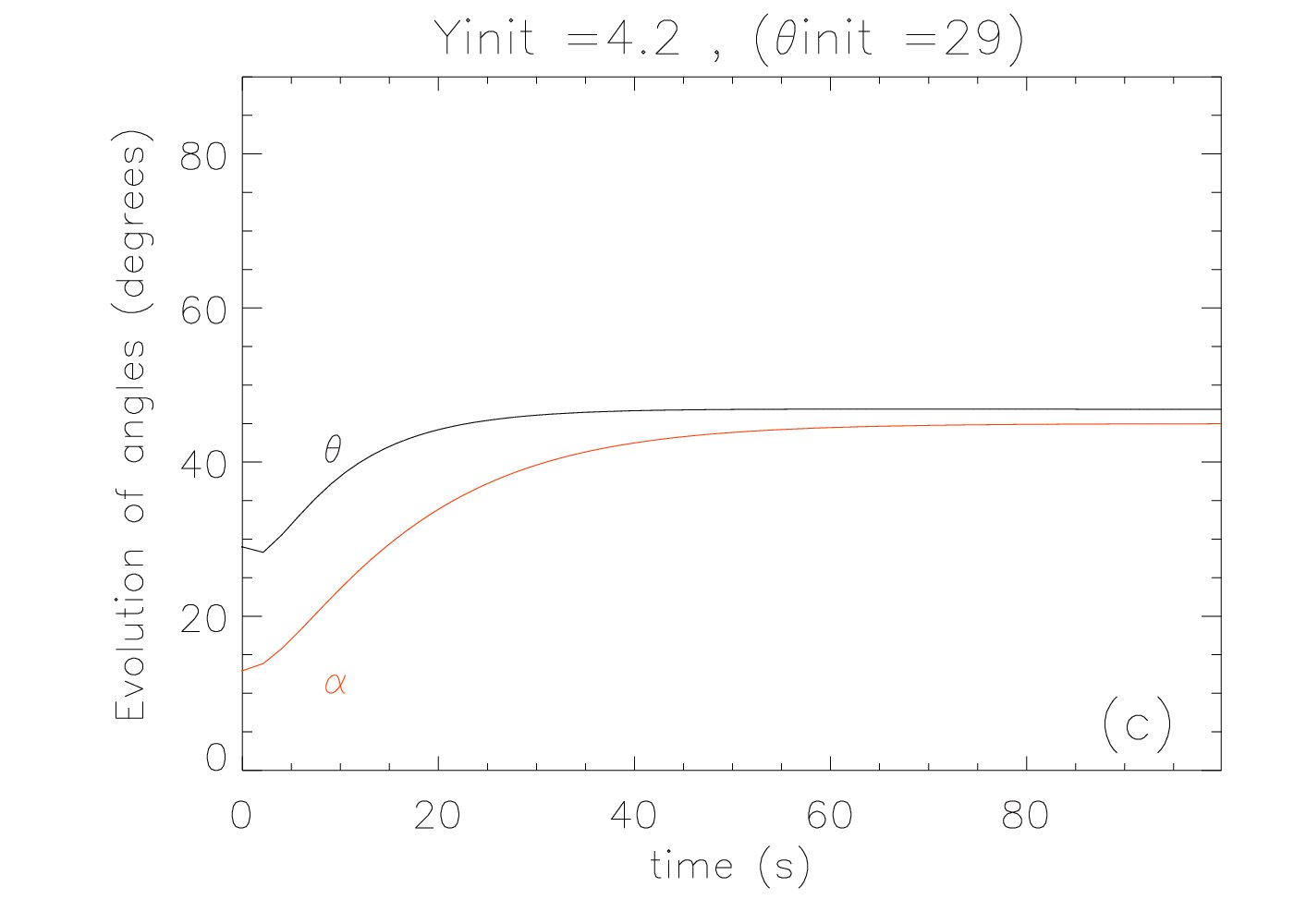}

   \caption{{Same quantities as in Fig. \ref{theta_alpha_simple1_0.1}, but for model 3. For a detailed discussion, see main text.}}
 \label{theta_alpha_simple3_0.1}
\end{figure*}
 \begin{figure*}
%   \begin{center}
\centerline{
   \includegraphics[width=6cm]{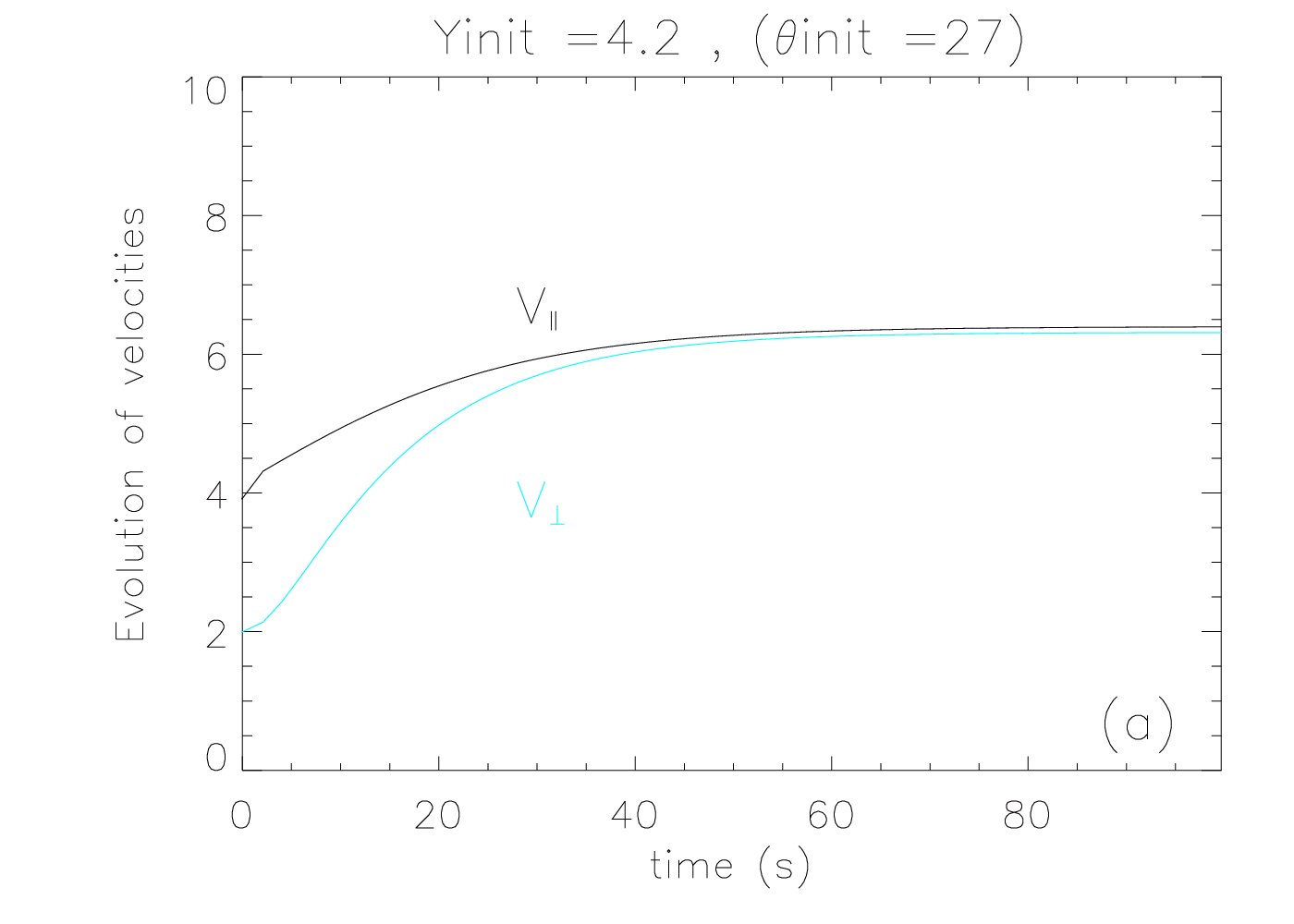}
  \includegraphics[width=6cm]{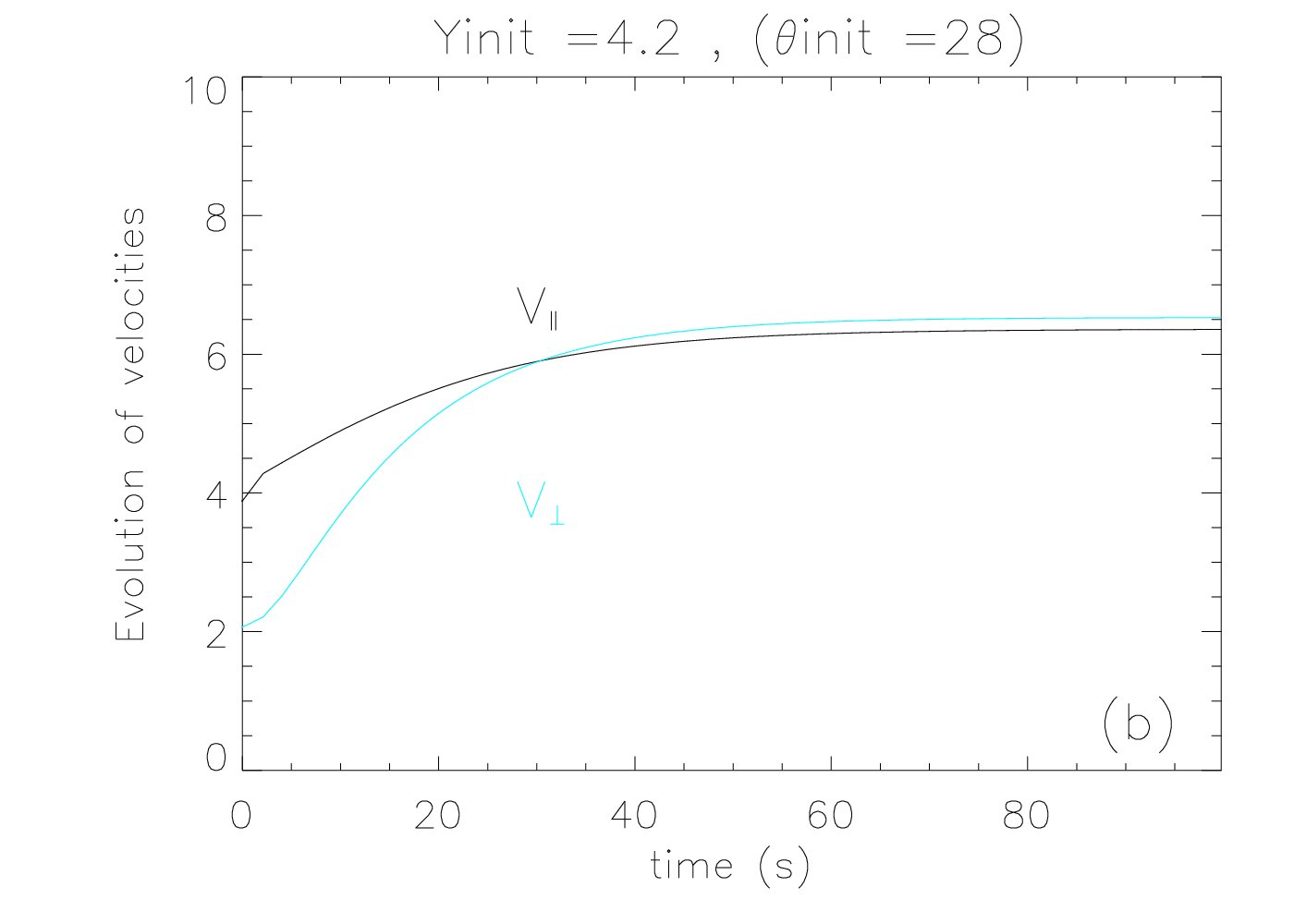}
  \includegraphics[width=6cm]{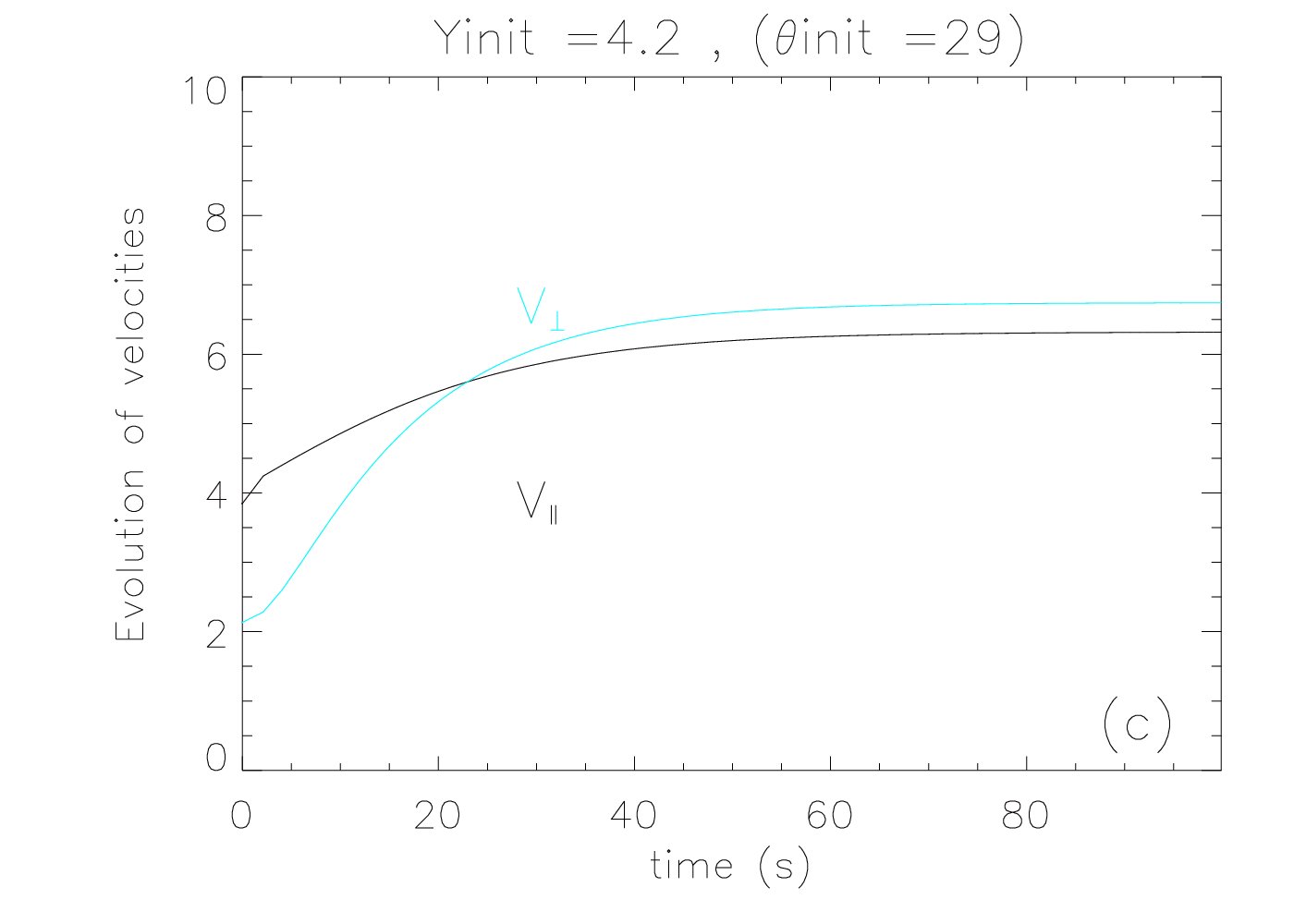}}

   \caption{{Same quantities as in Fig. \ref{vpar_vperp_simple1_0.1}, but for model 3. For a detailed 
discussion, see main text.}}
   \label{vpar_vperp_simple3_0.1_velocities}
\end{figure*}

Again, we show the results for three different values of the initial pitch angle in Figs. \ref{theta_alpha_simple3_0.1} and \ref{vpar_vperp_simple3_0.1_velocities}. 
The three values ($\theta_i=27^\circ$, $28^\circ$, and $29^\circ$) have been chosen such that they are in the region of values around the critical initial
pitch angle, which is very close to $26^\circ$.
For this case,
due to the modification of the time dependence of $B(t)$ there is an initial decrease in the pitch angle before it starts to increase in the later stages
of the evolution. 
{This is similar to what was found in model 1 for a later time. In model 2, the time evolution of the pitch angle differs because the magnetic field
strength $B(t)$ increases exponentially from $t=0$ on.\footnote{It is possible to generate plots for cases with initially decreasing pitch angle as well, if
one increases the $B$-field timescale $\tau$ considerably. We do not show plots of such cases here in order to keep the paper concise.}
Model 3 again shows very similar qualitative features to our kinematic MHD model results and to the other two simplified
model cases.}

All of the three simplified model cases have qualitatively similar features and those features are also qualitatively similar to what we found from the
guiding centre orbit calculations in Sect. \ref{sec:pitch}. In particular, these findings seem to corroborate that there is a critical initial pitch angle
below which particle orbits escape and above which particle orbits are trapped in a CMT. This critical initial pitch angle has a value, which is greater than the initial loss cone angle, but smaller than the asymptotic loss cone angle, and may vary from field line to field line.
We have also confirmed that while Fermi acceleration may be dominating in the initial phases of a particle orbit, betatron acceleration will
take over at some point and thus generally the pitch angle will increase with time. Fermi acceleration and betatron acceleration in
realistic CMTs will occur simultaneously and cannot be treated independently.

\section{Summary and conclusions} 

\label{sec:conclusions}

We have presented a detailed investigation of the conditions that affect the trapping and escape of particle orbits in models of collapsing magnetic traps, starting
with the investigation of guiding centre orbit calculations in the kinematic MHD CMT model of \citet{Giul2005}. Based on the results of our investigations
and on observations made previously by \citet{Giul2005} and \citet{Keith_grady_2012}, we designed a simplified schematic model for CMT particle acceleration
and studied three different implementations of this schematic model, which all qualitatively corroborated the previous findings.

We showed that for each magnetic field line in a collapsing magnetic trap there is a critical initial pitch angle, which divides particle orbits into trapped orbits and
escaping orbits. This critical initial pitch angle is greater than the initial loss cone angle for the field line, but smaller than the value of the asymptotic loss cone angle 
for the field line as $t\to \infty$.{ We also investigated whether the critical initial pitch angle depends on the initial energy of the particle, and we found 
that it does not.}

For orbits with initial pitch angle close to the critical value, Fermi acceleration can dominate in the initial phases, but betatron acceleration will become
dominant at later times. In the periods where Fermi acceleration dominates over betatron acceleration, the pitch angle will decrease and when
betatron acceleration dominates the pitch angle will increase. Due to the nature of CMTs, both mechanisms will always operate simultaneously
\citep[as already stated by][]{Kar_Barta_2006}, but the efficiency of Fermi acceleration has to decrease on a particular field line during the time
evolution of a CMT because the motion of the field line must slow down. On the other hand, the magnetic field strength can still continue to increase
due to the pile-up of magnetic flux from above.

In the present paper, we have for simplicity excluded the possibility of either Coulomb collisions, wave-particle interactions, or turbulence on the trapping or escape of 
particles from CMTs. Obviously, each of these mechanisms may change the results we have found in this  paper. 
The effect of Coulomb collisions on particle acceleration in simple CMT models has been considered by \citet{Kova_somov_2003}, \citet{Kar_kos_2004}, and \citet{Boga_somov_2009}. \citet{Mino_2011} have investigated the effects of pitch angle scattering by Coulomb collisions on trapping and escape, but
without considering the effect of dynamical friction. \citet{Kar_Barta_2006} included the effects of Coulomb collisions in their MHD model of a CMT. We also
mention the recent paper by \citet{winter_2011}, although a static field is used in that paper.

Obviously, wave-particle interaction and/or turbulence added to CMT model would also change both the energisation process and the pitch angle evolution
of particle orbits in a collapsing magnetic trap. As already mentioned by \citet{Keith_grady_2012}, a combination of a CMT model with stochastic acceleration
mechanisms \citep[see, e.g.,][]{miller} could provide a link between those acceleration mechanisms and the standard flare scenario along the lines proposed by e.g., \citet{hamilton:petrosian92},
\citet{park:petrosian95}, \citet{petrosian:liu04}, and \citet{liu:etal08}.

\begin{acknowledgements}
The authors would like to thank the referee for very useful comments and acknowledge useful discussions with Eduard Kontar, Peter Cargill, Marian Karlicky, and Mykola Gordovskyy.
This work was financially supported by the UK's Science and Technology Facilities Council.
\end{acknowledgements}

%%%%%%%%%%%%%%%%%%%%%%%%%%%
%start of bibliography
%%%%%%%%%%%%%%%%%%%%%%%%%%%%%%
\bibliographystyle{aa}
\bibliography{paper}

\appendix
\section{Simplified model: mathematical description}
\label{appendix}

We assume here that the functions $v_{LT}(t)$ and $B(t)$ are known. In the coordinate system used, $v_{LT}(t)$ should be negative as the
height of the field line apex should decrease in time. $B(t)$ should be a function that increases with time. For both functions, obviously many
different choices are possible and we therefore imposed the additional condition of simplicity on the functions.

As discussed in the main text, we make the simplifying assumption that the mirror height of each particle orbit is fixed. Without a detailed magnetic field model
and an explicit calculation of the corresponding particle orbits (which is exactly what we want to avoid in our simplified model), we can only choose
an ad hoc position of the mirror height. We know, however, that the mirror height should be coupled with the value of the initial pitch angle of a particle orbit and that
the mirror height should decrease with a decrease of the initial pitch angle, with the mirror height eventually reaching zero (lower boundary) when 
the initial pitch angle $\theta_{init}$
is equal to the initial loss cone angle $\alpha_{init}$. This relation can be parametrized as:
\begin{equation}
y_{mirror} = y_{init}\left[ 1 - \frac{f(\theta_{init})}{f(\alpha_{init})}\right],
\label{mirrorheight}
\end{equation}
where $f(\theta)$ is a function, which monotonically increases from a value of $0$ at $\theta = 90^\circ$. 
In this paper, we have chosen $f(\theta) = \cot^q \theta$, with $q>0$ a real parameter, but other choices are, of course, also possible.

The simplified model can be completely expressed in terms of motion in the $y$-direction and all other quantities can be derived from that. We start
by imagining the state of the system just after the particle has bounced off the field line apex for the $i^{th}$ time. Let the time of the
bounce be $t_i$ and the position $y_i = y(t_i)$. The velocity of the particle orbit in the 
$y$-direction then has the value $v_{y,i}$. The particle will then move down to the mirror point with this velocity, bounce off the mirror point
without changing the absolute value of its velocity (it is at this point that the assumption of the mirror points being static simplifies matters enormously) and
move back up to the field line apex for the next bounce off the loop top at time $t_{i+1}$. During that bounce the $v_y$-component of the velocity will change
 to
\begin{equation} 
 v_{y,i+1} = v_{y,i} + 2 v_{LT}(t_{i+1}).
 \end{equation}
 The particle will have travelled a total distance of
$ y_{i+1} + y_i - 2 y_{mirror}$ (distance down plus distance up) 
in the time $t_{i+1} - t_i$, leading to the equation
\begin{equation}
y_{i+1} = - y_i +  2 y_{mirror} + (t_{i+1} - t_i) v_{y,i}.
\label{yequation1}
\end{equation}
On the other hand, because the field line apex is moving with velocity $v_{LT}(t)$, we also have
\begin{equation}
y_{i+1} = y_i + \int\limits_{t_i}^{t_{i+1}} v_{LT}(t) \, dt = y_i + F_y(t_{i+1}) - F_y(t_i)
\label{yequation2}
\end{equation}
with $F_y(t) = \int v_{LT} (t) dt$. We would like to point out that the time $t_{i+1}$ of the next bounce is unknown and has to be calculated.

This can be done by combining Eqs. (\ref{yequation1}) and (\ref{yequation2}) to eliminate the equally unknown $y_{i+1}$ to get
the transcendental equation
\begin{equation}
F_y(t_{i+1}) - v_{y,i} t_{i+1} = F_y(t_i) - v_{y,i} t_i - 2 (y_i - y_{mirror}).
\label{ti+1equation}
\end{equation}
 Usually Eq. (\ref{ti+1equation}) will have to be solved numerically for $t_{i+1}$ using an iterative scheme, such as e.g., the Newton-Raphson method. However,
 for the case of a constant $v_{LT}(t)$ the equation becomes linear and can be solved analytically. 
 
 Once $t_{i+1}$ is known, all other relevant quantities can be determined and the process can be repeated until the end of the calculation.

\end{document}